\documentclass[fleqn,usenatbib]{mnras}
\usepackage[table,xcdraw]{xcolor}
\usepackage{newtxtext,newtxmath}
\usepackage{graphicx}
\usepackage{subfigure}
\usepackage[T1]{fontenc}
\usepackage{graphicx}	
\usepackage{caption}
\usepackage{amsmath}
\usepackage{hyperref}
\usepackage{rotating}
\hypersetup{colorlinks,citecolor=blue,filecolor=blue,linkcolor=blue,urlcolor=blue}

\title[A comparative study of NLSy1 and BLSy1 galaxies.]{ A comparative study of the physical properties for a representative sample of Narrow and Broad-line Seyfert galaxies}

\author[Vivek Kumar Jha et al.]{
Vivek Kumar Jha\thanks{E-mail: vivekjha.aries@gmail.com}$^{1,2}$,
Hum Chand,$^{1,3}$
Vineet Ojha,$^{1,4}$
Amitesh Omar,$^{1}$ and 
Shantanu Rastogi$^{2}$
\\
$^{1}$Aryabhatta Research Institute of observational sciencES (ARIES), Nainital, \it {263002}; India\\
$^{2}$Department of Physics, Deen Dayal Upadhyaya Gorakhpur University, Gorakhpur, \it {273009}; India\\
$^{3}$Department of Physics \& Astronomical sciences, Central University of Himachal Pradesh, Dharamshala, \it{176215}; India\\
$^{4}$Physical Research Laboratory (PRL), Astronomy and Astrophysics Division, Ahmedabad, {\it 380 009}; India}

\date{Accepted XXX. Received YYY; in original form ZZZ}
\pubyear{2021}

\begin{document} 
\label{firstpage}
\pagerange{\pageref{firstpage}--\pageref{lastpage}}
\maketitle
\begin{abstract}
We present a comparative study of the physical properties of a homogeneous sample of 144 Narrow line Seyfert 1 (NLSy1) and 117 Broad-line Seyfert 1 (BLSy1) galaxies. These two samples are in a similar luminosity and redshift range and have optical spectra available in the 16$^{th}$ data release of Sloan Digital Sky Survey (SDSS-DR16) and X-ray spectra in either XMM-NEWTON or ROSAT. Direct correlation analysis and a Principal Component Analysis (PCA) have been performed using ten observational and physical parameters obtained by fitting the optical spectra and the soft X-ray photon indices as another parameter. We confirm that the established correlations for the general quasar population hold for both types of galaxies in this sample despite significant differences in the physical properties. We characterize the sample also using the line shape parameters, namely the asymmetry and kurtosis indices. We find that the fraction of NLSy1 galaxies showing outflow signatures, characterized by blue asymmetries, is higher by a factor of about 3 compared to the corresponding fraction in BLSy1 galaxies. The presence of high iron content in the broad-line region of  NLSy1 galaxies in conjunction with higher Eddington ratios can be the possible reason behind this phenomenon. We also explore the possibility of using asymmetry in the emission lines as a tracer of outflows in the inner regions of Active Galactic Nuclei. The PCA results point to the NLSy1 and BLSy1 galaxies occupying different parameter spaces, which challenges the notion that NLSy1 galaxies are a subclass of BLSy1 galaxies.

\end{abstract}


\section{INTRODUCTION} \label{section:intro}

Seyfert-1 type galaxies have a lower luminosity nucleus as compared to quasars. These Active Galactic Nuclei (AGN) exhibit both broad and forbidden narrow emission lines in their spectra. The Seyfert class of AGN has been broadly divided into various subcategories based on the strength or the Full Width at Half Maximum (FWHM) of the Balmer $H\beta$ emission line \citep[e.g., see][for a review]{Netzer2015}. Narrow-line Seyfert 1 galaxies (NLSy1 galaxies) are a special subclass of AGN having narrower broad Balmer line widths with FWHM of broad $H\beta$ emission line $< 2000$ km s$^{-1}$ \citep{Goodrich1989}, a small flux ratio of the [O III] $\lambda$5007 to $H\beta$ line ($[O III]/H\beta < 3)$ \citep{Osterbrock1985}, stronger optical Fe II  emissions. These AGN show usually steeper soft X-ray spectra and rapid X-ray and sometimes optical flux variability \citep[see][and references therein]{Rakshit2017, Ojha2021}.  It is assumed that these properties are due to the central supermassive black hole (SMBH) being less massive but accreting at a very high rate \citep{Sulentic2000, Boroson2004}. It has been proposed that the NLSy1 galaxies are {\it younger} versions of Broad-line Seyfert 1 galaxies (BLSy1)  galaxies only and can be assumed to be in different evolutionary stages \citep[see][]{Mathur2000, Williams2018}. They have been known to have higher accretion rates as compared to the BLSy1  galaxies, which makes it one of the defining parameters in the classification \citep{Pounds1994, Collin2004, Sulentic2000}. 

The prominent emission lines arising from the Broad Line Region (BLR) are broadened to line widths of thousands of km s$^{-1}$ due to gas in virial motion around the central SMBH \citep{Gaskell2000, Peterson2004}. Direct observation of the BLR is desired in order to understand the dynamics of gas in these regions, but it has been a difficult challenge, and only recently \citet{Sturm2018, Gravityb, Gravityc} have spatially resolved the inner regions of a few AGN. Reverberation mapping (RM), which works in the time domain instead of the spatial domain, has been a method of choice to explore the dynamics of matter in the BLR \citep[see][]{Bahcall1972, Blandford1982, Peterson2004, Bentz2010, Grier2012, rosa2015, Du2016, Du2018, Lira2018, Wang2019}.

The RM measurements have yielded a strong relationship between the size of the BLR ($R_{BLR}$) and the optical luminosity measured at 5100 \AA{} ($L_{5100}$) \citep{kaspi2000, Bentz2009}. The velocity resolved reverberation mapping has enabled a more comprehensive understanding of these regions with the construction of velocity delay maps \citep{Grier2013, Xiao2018}. The dynamical modeling of a few AGN suggests that the gas in the BLR is dominated by Keplerian motion with traces of outflow and inflow \citep{Pancoast2014}. However, the origin and the geometry of the BLR  so far remain a mystery, and several models have been explored and debated for a very long time. For instance, the disk wind model \citep[][]{Gaskell2000, Gaskell2013, Baskin2018, Matthews2020} has been successful in explaining the outflowing profiles seen in the CIV emission line \citep{Gaskell2016}. The failed radiation-driven winds model proposed by \citet{Czerny2017} states that the material rising from the disk is expected to emit the low ionization lines such as Mg~{\sc ii} and H$\beta$ emission lines. This material spends more time in the rising phase than in the falling phase, hence the effect might appear like an outflow, although no actual outflow exists. Another model has been proposed by \citet{Wang2017} in which clumps of cloud from the dusty torus, tidally disrupted by the central SMBH, can give rise to the emission lines in the BLR.

While the known number of AGN extends into hundreds of thousands \citep{Veron-Cetty2000}, the reverberation mapping information is available for only approximately 120  AGN \citep[see][for a comprehensive database of reverberation mapped AGN]{Bentz2015}. Much of our understanding has relied on the statistical analysis of a sample of AGN constrained by various limits; to infer the physical properties of the inner regions of these AGN.  A remarkable work was done by \citet{Boroson1992}, where they performed a Principal Component Analysis (PCA) on the properties derived from X-ray, optical, and radio wavelength data for a set of 87 AGN. With this dataset, they derived that the Eigenvector 1 (E1), which is driven by the anti-correlation of the ratio of equivalent width (EW) of Fe~{\sc II} lines in the optical band and the FWHM of $H\beta$ emission line ($R_{fe}$) is the primary cause of variability in the parameters. \citet{Sulentic2000} and later subsequent works e.g., \citet{Zamfir2010} established the foundations of the four-dimensional EV1 (4DE1) formalism, which includes FWHM of $H\beta$ and $R_{fe}$ as two of the main components. These two quantities are related to the SMBH mass and the Eddington ratio and the ongoing results have yielded a so-called quasar main sequence \citep[see][and references therein]{shen2014, Sulentic2015, Marziani2018}.

One of the important consequences of such statistical studies on a set of AGN has been that the prominent broad emission lines in a few Type-1 AGN show asymmetric profiles \citep[see][]{Sulentic1989,corbin1995, Brotherton1996, Marziani1996, Zamfir2010, Negrete2018, Wolf2020}. The asymmetric emission line profiles raise essential questions in our understanding of the gas in the BLR itself, like whether the gas is in virial motion around the SMBH or is it in the form of an outflowing/inflowing motion related to the accretion disk. A red asymmetric emission profile may indicate the presence of inflowing gas rotating nearer to the SMBH, while the blue asymmetric emission profiles may indicate outflow being induced by the presence of a disk-wind structure \citep{Zamfir2010}. The asymmetry in the Balmer emission lines yields crucial information about the gas motion in the vicinity of the SMBH, and hence proper understanding of the cause of asymmetric profiles is imperative. While in the high ionization line like the C~{\sc iv} emission line, the blue ward asymmetry is explained by the two-component BLR model \citep[e.g., see][]{Gaskell2016, Sulentic2017}, a similar analogy is not used for low ionization lines such as the $H\beta$ emission line. The characterization of the AGN in terms of their emission line shapes and shifts and their correlation with observational and physical parameters such as the FWHM of $H\beta$ emission line and accretion rate has been demonstrated in \citet{Zamfir2010} where they conclude that the AGN with $H\beta$ FWHM $\geq$ 4000 km s$^{-1}$ show different characteristics than the ones with the lower value of FWHM. 

\begin{figure}
\includegraphics[width=8cm,height=6cm]{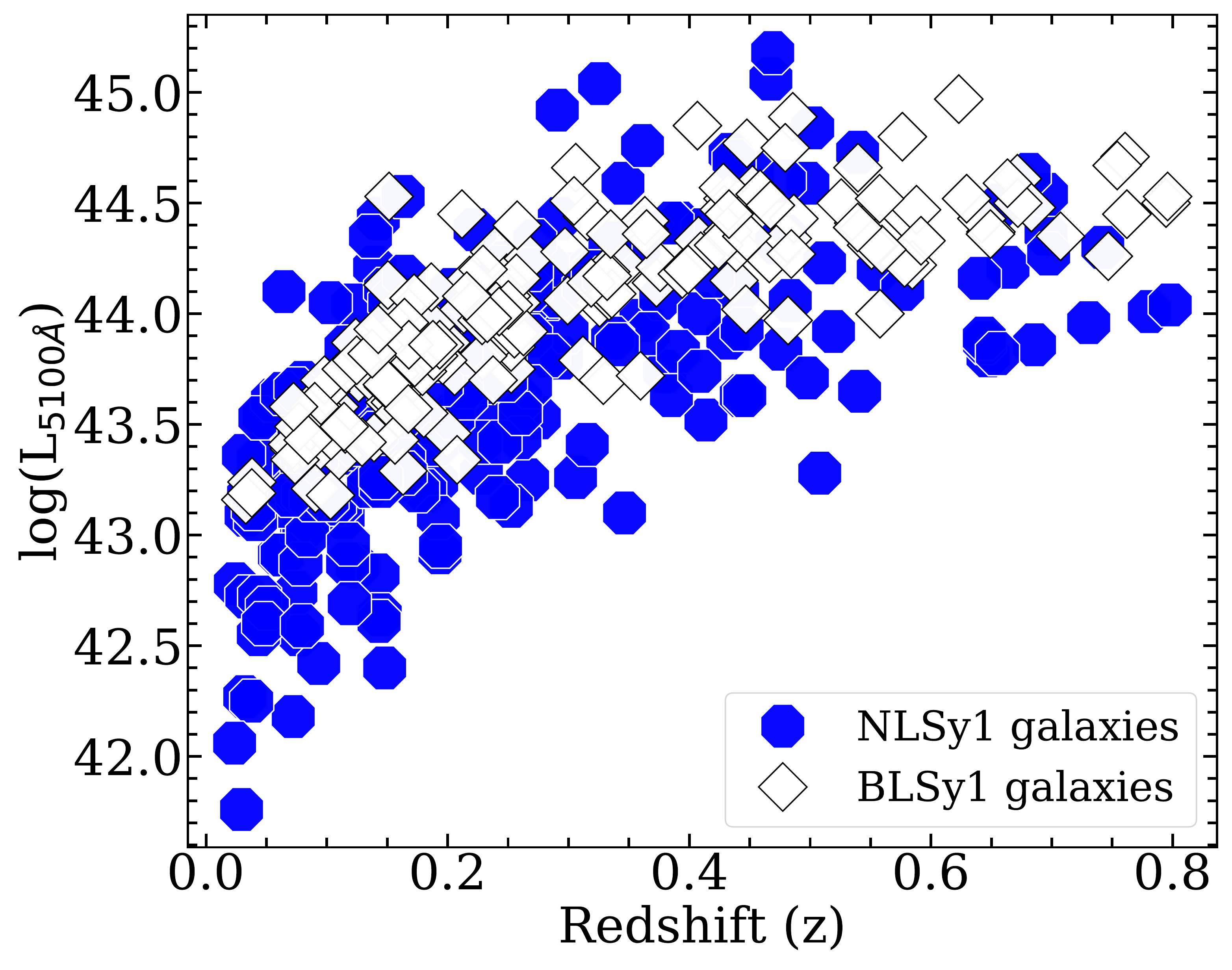}

\caption{The luminosity- redshift (L-z) distribution for the entire sample of 221 NLSy1 and 154 BLSy1  galaxies being used for this work. This sample is limited to a redshift of 0.8 due to the upper wavelength limit of $H\beta$ emission line in SDSS. The redshift values were obtained from the values available in the SDSS database while the $L_{5100}$ \AA{} values were obtained using the spectral fitting procedure outlined in Sec.~\ref{analysis}.}
\label{sample}
\end{figure}

In the type 1 AGN family, the NLSy1 galaxies are located at the extreme end of the population and appear as a class of AGN with high accretion rates, which may be a defining criterion as compared to the traditional FWHM based classification of these AGN \citep{Marziani2018p}. In the recent works, statistical study of the properties of NLSy1 galaxies with a control sample of BLSy1 galaxies is explored \citep[e.g., see][]{Nagao2001, zhou2006, zhou2010, Jin2012, Xu2012, Cracco2016, Ojha2020, Waddell2020}; however, the line shape parameters have not been covered for a large sample. It remains to be understood what fraction of the NLSy1 galaxies shows asymmetric emission line profiles and possible causes behind this phenomenon. It is essential to know how emission line asymmetries correlate with other observational and physical parameters and whether this behavior is peculiar compared to the general type 1 AGN population.  While high accretion rates in NLSy1 galaxies make them stand out from the general population of Seyfert galaxies, it becomes imperative to understand the variation in these properties for a representative population of these galaxies. A comparison of the properties of NLSy1 galaxies and a control sample of BLSy1 galaxies in the context of their BLR geometry characterized by the emission line features such as the asymmetry and kurtosis helps understand the relationship between the two types of AGN. This is also important to understand whether the NLSy1 galaxies form a particular subclass of the BLSy1 galaxies.

In this work, we have performed a statistical study to understand the observational and physical parameters responsible for driving the variations in the population of NLSy1 and BLSy1 galaxies. This paper is organized as follows: we describe the sample in Sec. \ref{sample-section}, which is followed by the analysis in Sec. \ref{analysis} where we present the fitting procedure and the estimation of physical parameters. The results obtained in this work are presented in Sec. \ref{results}. The discussion and interpretation of the results are presented in Sec. \ref{discussion}, and we conclude with a summary in Sec. \ref{conclusions}.

\begin{figure}
\includegraphics[width=8cm,height=6.5cm]{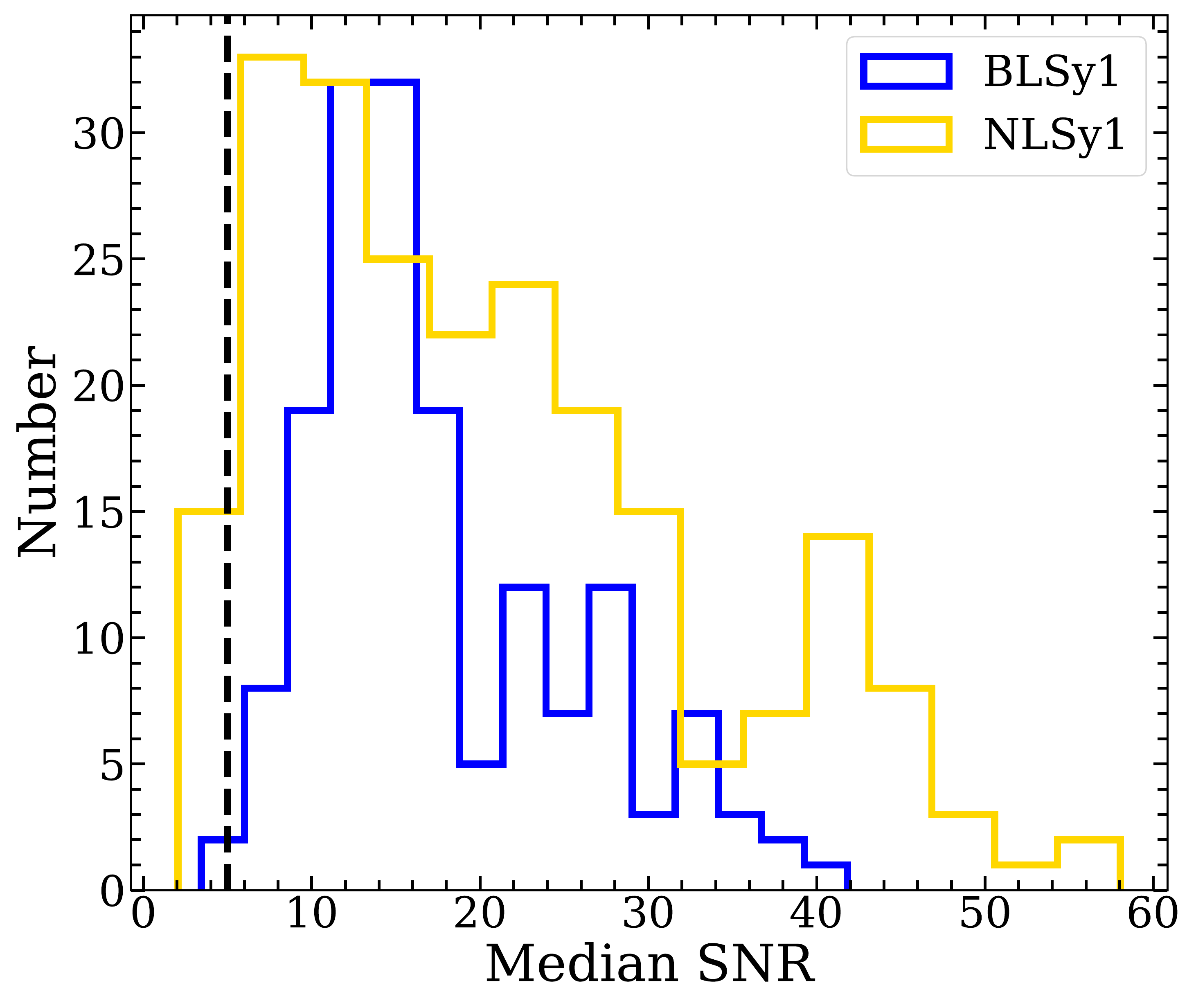}

\caption{Distribution of median Signal to Noise Ratio (SNR) for the parent sample of 221 NLSy1 and 154 BLSy1  galaxies obtained from \citep{Ojha2020}. The vertical dashed red line is drawn at SNR = 5, rejecting 15 NLSy1 galaxies and 1 BLSy1 galaxy from the parent sample, with SNR lower than this limit.}
\end{figure}

\section{Sample}
\label{sample-section}


Our present sample comprising of both NLSy1 galaxies and BLSy1  galaxies, matching in redshift (see Figure~\ref{sample}) has been drawn from~\citet{Ojha2020} where they have presented a detailed comparative study of a redshift-matched sample of 221 NLSy1 galaxies and 154 BLSy1  galaxies having optical observation in Sloan Digital Sky Survey (SDSS) and X-ray observation either in Röntgensatellit (ROSAT\footnote{\url{https://heasarc.gsfc.nasa.gov/docs/rosat/rosat.html}}) or in X-ray Multi-Mirror Mission (XMM\footnote{\url{https://www.cosmos.esa.int/web/xmm-newton}}).  Firstly, we obtained the optical spectrum of all the sources from the sample within a position offset of 3 arcseconds from the recent data release 16 of SDSS (SDSS-DR16)\footnote{\url{https://dr16.sdss.org/}}. More information about the SDSS database of spectra is available in \citet[][]{York2000, Abazajian2009,Shen2011,Rakshit2017,Lyke2020}. Here, we avoided the repetition of any source with multi-epoch spectra by retaining only the epoch with the highest signal-to-noise ratio (SNR). Furthermore, to get precise emission line detection which is important for fitting the spectra, we put a minimum limit of SNR $\geq$ 5. This criterion was satisfied by 206 NLSy1 galaxies and 153 BLSy1  galaxies. The single epoch optical spectrum of each source was analyzed using the publicly available code PyQSOFIT\footnote{\url{https://github.com/legolason/PyQSOFit}}.  The spectra were dereddened using the dust reddening maps available in \citet{Schlegel1998}\footnote{\url{https://github.com/kbarbary/sfdmap}} and brought to the rest frame using the redshift values available in the headers of the individual spectrum.

\begin{figure*}

\includegraphics[width=19cm,height=9cm]{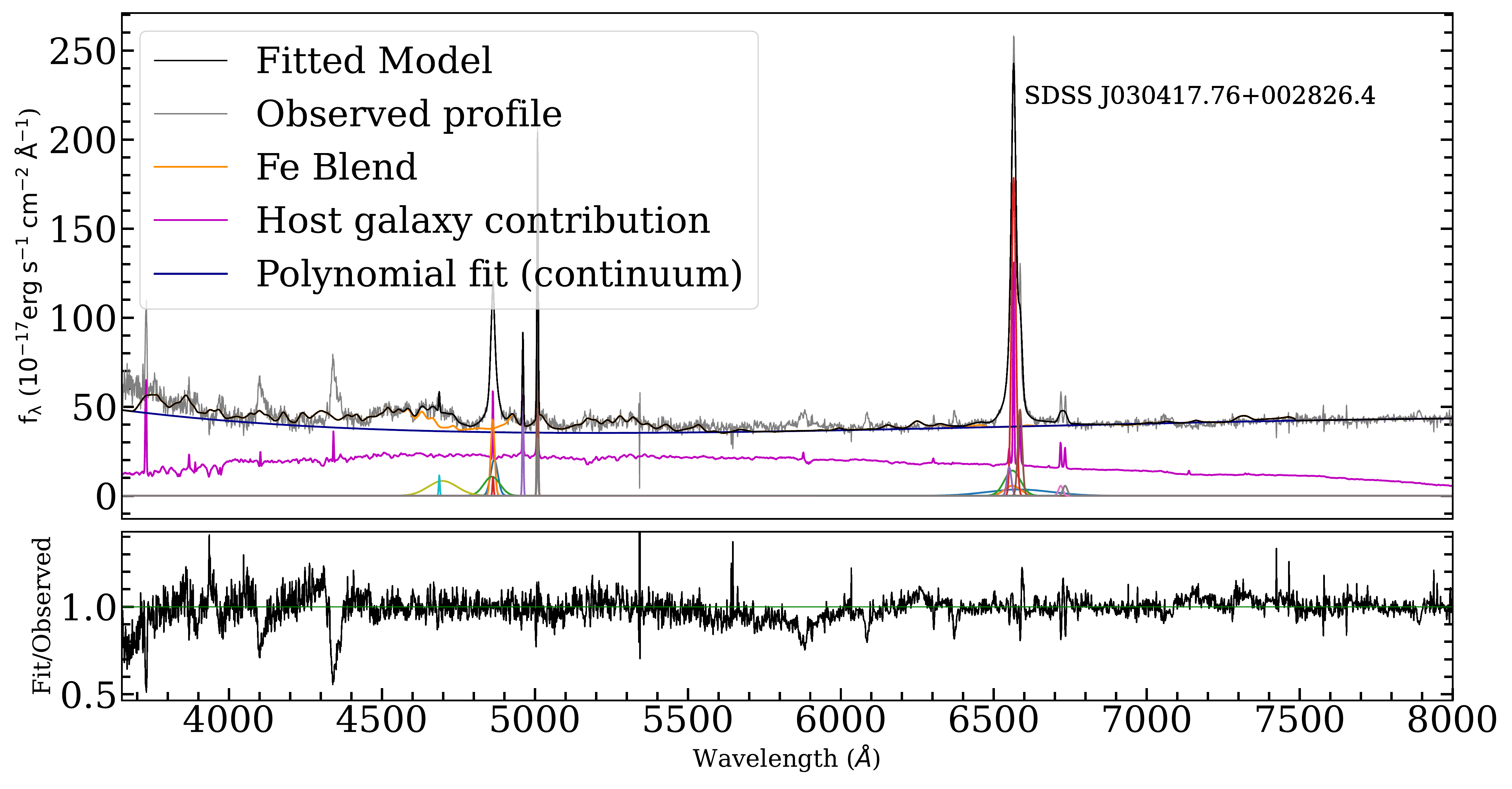}

\caption{Demonstration of the fitting procedure: The host galaxy has been decomposed using available templates in \citet{Yip2004}, and the Fe contamination has been removed using the templates available in \citet{Boroson1992}. The continuum has been fit using a power law, in order to fit the emission lines, a combination of up to 3 Gaussian profiles was used: a narrow component with FWHM $\leq$ 1200 km s$^{-1}$, a broad component with FWHM  $\leq$ 2300 km s$^{-1}$ for the NLSy1 galaxies and a very broad component $\geq$ 2300 km s$^{-1}$ in some cases. The limit of 2300 km s$^{-1}$ was removed in the case of BLSy1 galaxies. The [OIII] doublet is fitted using a single Gaussian component with its width tied to the $H\beta$ narrow component.}
\label{fitting_demo}
\end{figure*}

\begin{figure*}

\includegraphics[width=7cm,height=8cm]{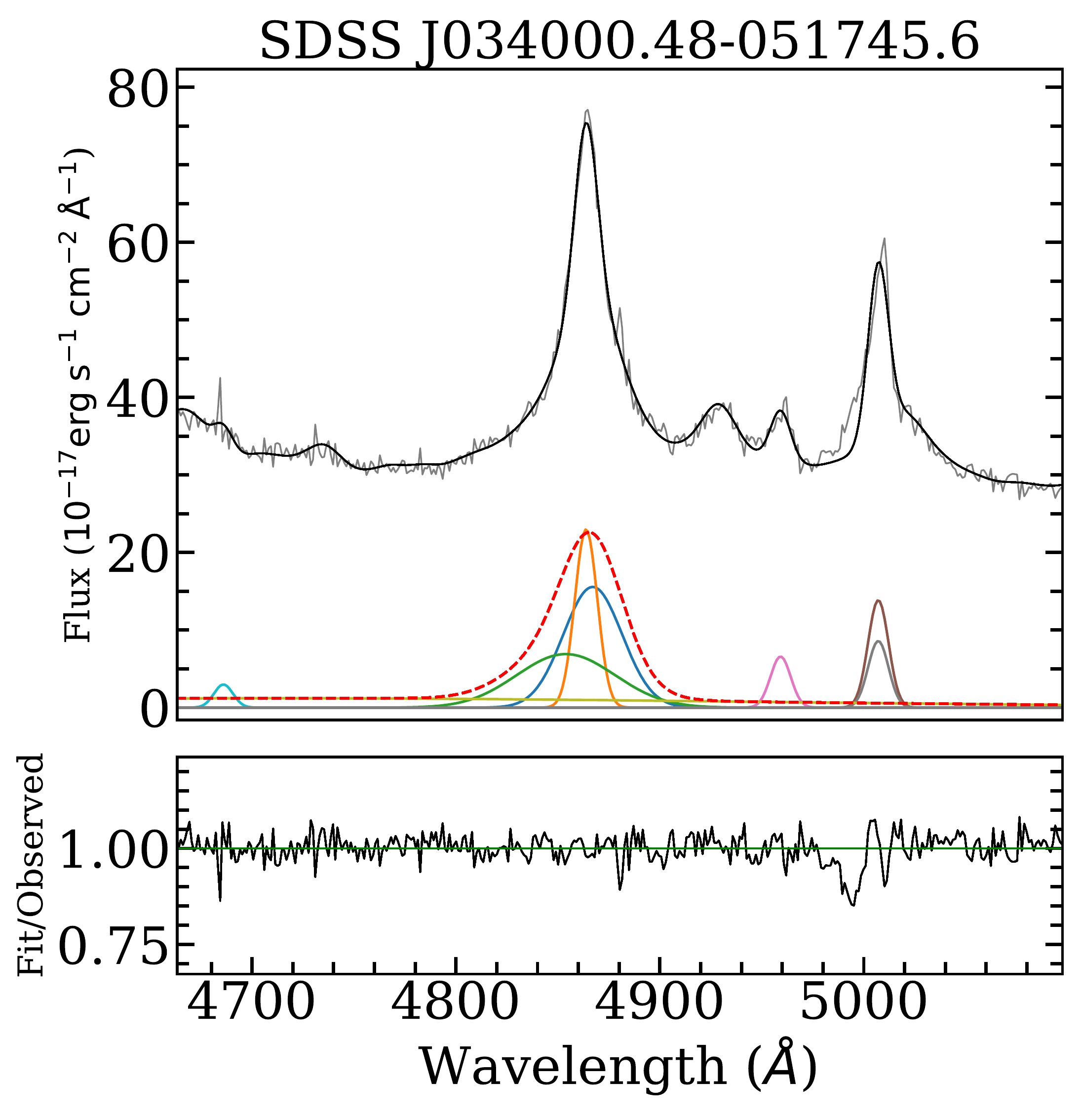}
\includegraphics[width=7cm,height=8cm]{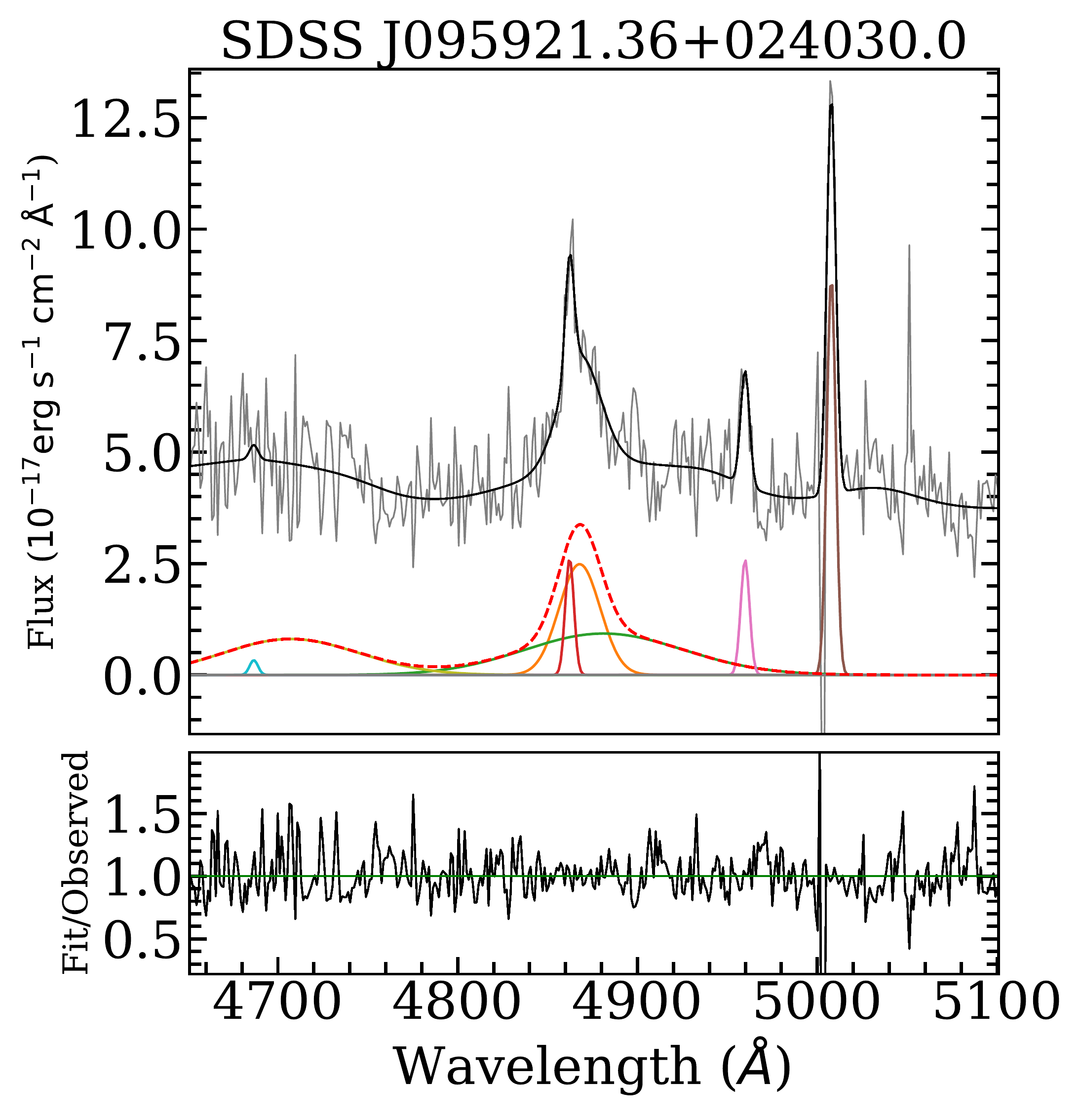}

\caption{Demonstration of asymmetric $H\beta$ emission profiles. The multiple Gaussian profiles used to fit the H$\beta$ emission line are also shown. The combination of broad components used to estimate AI is shown as a dashed red line. An example of a {\it blue} asymmetric profile having a negative asymmetry index of $-$ 0.152 is shown in panel (a), while the panel (b) shows an example of {\it red} asymmetric profile having a positive asymmetry index of $+$ 0.227. The AI values are calculated using Equation \ref{ai_eqn} by utilizing the flux values at 75\% and 25\%, respectively.}
\end{figure*}

\begin{sidewaystable*}
\vspace{-18cm}
\fontsize{7}{8}\selectfont
\centering

\caption{ The table of properties obtained for the sample of 144 NLSy1 and 117 BLSy1 galaxies.}
\hspace{0.8cm}
\begin{tabular}{lrrrrrrrrrccrrr}

\hline
SDSS Name\footnote{Obtained from SDSS SAS.}  & z  \footnote{Source redshift based on single epoch spectra available in SDSS DR16. }  & FWHM ($H\beta$ ) \footnote{The broad component of $H\beta$ emission profile in the units of km s$^{-1}$.} & EW($H\beta$) \footnote{Equivalent width of $H\beta$ emission line.}  & [OIII]/$H\beta$ \footnote{ [OIII]/$H\beta$ emission line flux ratio.} & $H\alpha/H\beta$ \footnote{$H\alpha/H\beta$ broad component flux ratio.}& $R_{fe}$ \footnote{$R_{fe}$, calculated as the flux ratio of area covered by the broad Fe line between 4433 \AA{} and 4684 \AA{} and the flux of the $H\beta$ emission line.}  & log($L_{5100\AA})$ \footnote{ Optical luminosity at 5100\AA{}($\lambda$L$_{5100\AA}$ in units of ergs/sec/\AA}.   & $M_{BH}$ \footnote{SMBH mass in the units of log($M_{BH}/M\odot$)  obtained using the Radius-Luminosity relation and FWHM of $H\beta$ emission line.}   & log($R_{\textsc{edd}}$) \footnote{ Eddington ratio, the ratio of bolometric to Eddington luminosity.}  & AI \footnote{Asymmetry index for the H$\beta$ emission line calculated using Equation \ref{ai_eqn}(a).}    & KI \footnote{Kurtosis Index for the H$\beta$ emission line calculated using Equation \ref{ai_eqn}(b).}  &   $\Gamma_{X}$\footnote{The soft X-ray photon index calculated between 0.2-2 KeV energy range taken from \citet{Ojha2020}.}      \\

\hline
\multicolumn{12}{c}{NLSy1 galaxies}\\

\hline

J001137.20+144160.0 & 0.1319 & 705  $\pm$ 152       & 26.87  $\pm$ 7.09    & 0.36                           $\pm$ 0.13      & 4.65  $\pm$ 0.31    & 1.35  $\pm$ 0.50     & 43.853 $\pm$ 0.002      & 6.64     $\pm$ 0.30           & 0.67   $\pm$ 0.29      & 0.404  $\pm$ 0.174   & 0.114 $\pm$ 0.025  & 2.52        $\pm$ 0.18        \\
J003859.28+005450.4 & 0.4137 & 1537 $\pm$ 183       & 70.08  $\pm$ 14.72   & 0.34                           $\pm$ 0.10      & 1.79  $\pm$ 0.25    & 0.50  $\pm$ 0.15     & 43.915 $\pm$ 0.005      & 7.32     $\pm$ 0.17           & 0.03   $\pm$ 0.01      & 0.003  $\pm$ 0.001   & 0.321 $\pm$ 0.038  & 2.53        $\pm$ 0.21        \\
J011911.52-104532.4 & 0.1253 & 1478 $\pm$ 297       & 40.28  $\pm$ 13.16   & 0.00                           $\pm$ 0.00      & 0.87  $\pm$ 0.22    & 0.63  $\pm$ 0.29     & 43.241 $\pm$ 0.006      & 7.26     $\pm$ 2.84           & -0.27  $\pm$ 1.10      & 0.129  $\pm$ 0.517   & 0.265 $\pm$ 0.534  & 2.97        $\pm$ 0.66        \\
J014644.88-004044.4 & 0.0824 & 1699 $\pm$ 184       & 44.76  $\pm$ 7.23    & 0.53                           $\pm$ 0.12      & 2.85  $\pm$ 0.19    & 1.32  $\pm$ 0.30     & 43.545 $\pm$ 0.002      & 7.39     $\pm$ 0.15           & -0.24  $\pm$ 0.05      & -0.014 $\pm$ 0.003   & 0.316 $\pm$ 0.034  & 2.92        $\pm$ 0.31        \\
J014904.56+125746.8 & 0.7305 & 1454 $\pm$ 336       & 153.93 $\pm$ 44.27   & 0.00                           $\pm$ 0.00      & 0.00  $\pm$ 0.00    & 0.09  $\pm$ 0.04     & 44.583 $\pm$ 0.015      & 7.43     $\pm$ 0.33           & 0.41   $\pm$ 0.20      & 0.229  $\pm$ 0.106   & 0.345 $\pm$ 0.080  & 2.12        $\pm$ 0.47        \\
J020337.20-051406.0 & 0.5193 & 1746 $\pm$ 242       & 74.09  $\pm$ 4.49    & 0.00                           $\pm$ 0.00      & 2.03  $\pm$ 0.14    & 0.66  $\pm$ 0.06     & 44.160 $\pm$ 0.003      & 7.47     $\pm$ 0.20           & 0.04   $\pm$ 0.01      & 0.014  $\pm$ 0.004   & 0.228 $\pm$ 0.032  & 1.58        $\pm$ 0.58        \\
J020853.28-043354.0 & 0.5563 & 1513 $\pm$ 439       & 56.40  $\pm$ 8.48    & 0.28                           $\pm$ 0.06      & 2.15  $\pm$ 0.32    & 0.96  $\pm$ 0.20     & 44.345 $\pm$ 0.002      & 7.38     $\pm$ 0.41           & 0.26   $\pm$ 0.15      & 0.040  $\pm$ 0.023   & 0.296 $\pm$ 0.086  & 2.84        $\pm$ 0.84        \\
J021329.28-051138.4 & 0.443  & 705  $\pm$ 254       & 8.39   $\pm$ 3.24    & 3.22                           $\pm$ 1.75      & 6.93  $\pm$ 3.90    & 3.54  $\pm$ 1.93     & 44.080 $\pm$ 0.006      & 6.67     $\pm$ 0.51           & 0.79   $\pm$ 0.57      & 0.023  $\pm$ 0.017   & 0.347 $\pm$ 0.125  & 2.21        $\pm$ 0.80        \\
J021803.60-004337.2 & 0.3796 & 968  $\pm$ 506       & 68.57  $\pm$ 11.15   & 0.00                           $\pm$ 0.00      & 1.61  $\pm$ 0.16    & 0.75  $\pm$ 0.17     & 44.061 $\pm$ 0.005      & 6.94     $\pm$ 0.74           & 0.50   $\pm$ 0.53      & -0.148 $\pm$ 0.155   & 0.325 $\pm$ 0.170  & 2.84        $\pm$ 0.19        \\
J022452.32-040520.4 & 0.6952 & 859  $\pm$ 100       & 56.08  $\pm$ 12.07   & 0.39                           $\pm$ 0.12      & 0.00  $\pm$ 0.00    & 1.23  $\pm$ 0.38     & 44.589 $\pm$ 0.010      & 6.97     $\pm$ 0.16           & 0.87   $\pm$ 0.22      & -0.029 $\pm$ 0.007   & 0.299 $\pm$ 0.035  & 2.61        $\pm$ 0.27        \\
J022928.32-051124.0 & 0.3068 & 1338 $\pm$ 175       & 35.45  $\pm$ 2.62    & 0.42                           $\pm$ 0.04      & 4.46  $\pm$ 0.45    & 1.63  $\pm$ 0.17     & 44.573 $\pm$ 0.001      & 7.35     $\pm$ 0.19           & 0.48   $\pm$ 0.13      & -0.121 $\pm$ 0.032   & 0.313 $\pm$ 0.041  & 2.97        $\pm$ 0.23        \\

... &...  &... &... &... &... &... &... &... &... &... &... &...  \\
... &...  &... &... &... &... &... &... &... &... &... &... &...  \\

\hline

\multicolumn{12}{c}{BLSy1  galaxies}\\

\hline

J002113.20-020115.6 & 0.7621 & 9759  $\pm$ 1578      & 116.29 $\pm$ 19.26   & 0.39                           $\pm$ 0.09      & 0.00  $\pm$ 0.00    & 0.30  $\pm$ 0.07     & 44.644 $\pm$ 0.010      & 9.11      $\pm$ 0.23           & -1.21  $\pm$ 0.41      & 0.548    $\pm$ 0.177   & 0.568  $\pm$ 0.092  & 1.83        $\pm$ 0.35        \\
J011254.96+000314.4 & 0.2389 & 6561  $\pm$ 347       & 76.43  $\pm$ 13.97   & 0.14                           $\pm$ 0.04      & 3.29  $\pm$ 0.17    & 0.71  $\pm$ 0.18     & 44.397 $\pm$ 0.002      & 8.67      $\pm$ 0.07           & -0.99  $\pm$ 0.11      & 0.034    $\pm$ 0.004   & 0.290  $\pm$ 0.015  & 2.60        $\pm$ 0.24        \\
J012254.72+010108.4 & 0.1994 & 3506  $\pm$ 469       & 23.90  $\pm$ 5.20    & 0.36                           $\pm$ 0.11      & 2.94  $\pm$ 0.27    & 0.73  $\pm$ 0.22     & 43.548 $\pm$ 0.003      & 8.02      $\pm$ 0.19           & -0.87  $\pm$ 0.24      & -0.048   $\pm$ 0.013   & 0.299  $\pm$ 0.040  & 1.79        $\pm$ 0.38        \\
J013517.52-001940.8 & 0.3119 & 7242  $\pm$ 471       & 36.79  $\pm$ 1.02    & 0.31                           $\pm$ 0.01      & 4.04  $\pm$ 0.13    & 0.19  $\pm$ 0.01     & 43.931 $\pm$ 0.003      & 8.67      $\pm$ 0.09           & -1.31  $\pm$ 0.18      & 0.012    $\pm$ 0.001   & 0.340  $\pm$ 0.022  & 2.31        $\pm$ 0.30        \\
J014959.28+125656.4 & 0.432  & 4076  $\pm$ 275       & 87.32  $\pm$ 23.20   & 0.19                           $\pm$ 0.07      & 0.00  $\pm$ 0.00    & 0.80  $\pm$ 0.30     & 44.670 $\pm$ 0.003      & 8.36      $\pm$ 0.10           & -0.44  $\pm$ 0.06      & 0.095    $\pm$ 0.013   & 0.311  $\pm$ 0.021  & 2.24        $\pm$ 0.19        \\
J020744.16-060957.6 & 0.6496 & 9877  $\pm$ 699       & 102.77 $\pm$ 9.19    & 0.26                           $\pm$ 0.03      & 0.00  $\pm$ 0.00    & 0.44  $\pm$ 0.06     & 44.537 $\pm$ 0.003      & 9.07      $\pm$ 0.10           & -1.28  $\pm$ 0.19      & 0.477    $\pm$ 0.068   & 0.556  $\pm$ 0.039  & 2.07        $\pm$ 0.46        \\
J020840.56-062718.0 & 0.092  & 2509  $\pm$ 42        & 31.55  $\pm$ 4.61    & 0.00                           $\pm$ 0.00      & 3.57  $\pm$ 0.13    & 1.10  $\pm$ 0.23     & 43.449 $\pm$ 0.001      & 7.72      $\pm$ 0.02           & -0.63  $\pm$ 0.02      & -0.039   $\pm$ 0.001   & 0.315  $\pm$ 0.005  & 2.21        $\pm$ 0.11        \\
J021139.12-042606.0 & 0.4856 & 3099  $\pm$ 193       & 62.47  $\pm$ 2.69    & 0.03                           $\pm$ 0.00      & 2.51  $\pm$ 0.12    & 1.00  $\pm$ 0.06     & 45.025 $\pm$ 0.001      & 8.42      $\pm$ 0.09           & -0.02  $\pm$ 0.00      & 0.004    $\pm$ 0.001   & 0.284  $\pm$ 0.018  & 2.78        $\pm$ 0.19        \\
J021318.24+130643.2 & 0.4076 & 8659  $\pm$ 581       & 145.32 $\pm$ 17.39   & 0.17                           $\pm$ 0.03      & 0.00  $\pm$ 0.00    & 0.29  $\pm$ 0.05     & 44.527 $\pm$ 0.004      & 8.95      $\pm$ 0.10           & -1.17  $\pm$ 0.16      & 0.030    $\pm$ 0.004   & 0.300  $\pm$ 0.020  & 2.07        $\pm$ 0.24        \\
J021434.80-004243.2 & 0.4369 & 11843 $\pm$ 582       & 159.30 $\pm$ 28.22   & 0.53                           $\pm$ 0.13      & 0.00  $\pm$ 0.00    & 0.43  $\pm$ 0.11     & 44.210 $\pm$ 0.006      & 9.14      $\pm$ 0.07           & -1.60  $\pm$ 0.17      & -0.021   $\pm$ 0.002   & 0.306  $\pm$ 0.015  & 2.48        $\pm$ 0.14        \\
J021820.40-050426.4 & 0.6492 & 3440  $\pm$ 428       & 51.72  $\pm$ 10.87   & 0.30                           $\pm$ 0.09      & 0.00  $\pm$ 0.00    & 0.00  $\pm$ 0.00     & 44.586 $\pm$ 0.009      & 8.18      $\pm$ 0.18           & -0.34  $\pm$ 0.09      & -0.036   $\pm$ 0.009   & 0.312  $\pm$ 0.039  & 1.92        $\pm$ 0.26        \\

... &...  &... &... &... &... &... &... &... &... &... &... &... \\
... &...  &... &... &... &... &... &... &... &... &... &... &... \\
\hline
\end{tabular}

Note: Only a portion of the table is available here to show the form and content. The entire table is available in the online version.
\vspace{0.5cm}
\label{table1}
\end{sidewaystable*}

\section{Analysis}
\label{analysis}

\subsection{Spectral fitting}

 We used the publicly available code PyQSOFit \citep{Guo2018} to decompose the spectra into multiple components. Initially, the continuum model was prepared using the host galaxy components, the contribution from the iron line, and the accretion disk continuum, which reflects itself as a power-law component. A simple power-law component describes the accretion disk continuum satisfactorily \citep{Sexton2020}. We removed the host galaxy component using the Principal Component Analysis (PCA) method, obtained from the host galaxy templates available in \citet{Yip2004}. We used five components of the PCA in order to subtract the host galaxy emission. However, for many of the sources in our sample, the host galaxy decomposition could not be applied. At higher redshifts, the contribution from the host galaxy does not have a  large impact on the emission spectrum; hence we did not attempt to remove the host galaxy contribution from the spectra wherever it was not possible. The Fe blends were removed using the templates available in \citet{Boroson1992} which are available within the code itself. The final continuum model consisting of: the power-law component, the host galaxy template wherever applicable, and the Fe blends was subtracted from the original spectrum, which yielded the emission line components only. We were concerned with measuring the asymmetry in the $H\beta$ emission line; hence for the emission line fitting, we concentrated on the $H\beta$-[OIII] emission line complex. We used a combination of both narrow and broad Gaussian profiles to build the emission line model. We assumed the emission line complex to be composed of narrow and broad components representing the $H\beta$ emission from the Narrow line and Broad-line regions, respectively. The width of the narrow Gaussian components used for fitting the [OIII] doublet was tied with the narrow component of the $H\beta$ emission line, which physically indicates the emission coming from the same narrow-line region. This technique has been widely used while fitting the AGN spectra \citep[e.g., see][]{Chand2010,Rakshit2017, Sexton2020}. For fitting the broad profile of the $H\beta$ emission line, we used a combination of up to 3 Gaussian components. The broad central component was needed in all the cases, while additional Gaussian components were needed to fit the wings of the line in a few of the cases. We also allowed a very broad Gaussian component for a few sources. The existence of a very broad component has been proposed for a few AGN in the recent works   \citep[see][]{Gaskell2013, Sulentic2015, Marziani2018, Wolf2020} in which emission coming either from the innermost regions of the BLR or from the outer regions of the accretion disk results in a very broad component of the emission lines. \citet{Hao2005} set a criterion of FWHM larger than 1200 km s$^{-1}$ for defining broad line AGN. As a result, we set the limits for the width of the Gaussian profiles at 1200 km s$^{-1}$ for narrow [OIII] and $H\beta$ components, while  2300 km s$^{-1}$ was set for broad components and beyond 10000 km s$^{-1}$ for very broad components. The limit of 2300 km s$^{-1}$ was kept keeping in mind the previous works classifying the NLSy1 galaxies \citep[e.g., see][]{Rakshit2017}. For fitting the BLSy1  galaxies, we removed the upper limit of 2300 km s$^{-1}$ on the broad component while still allowing up to three Gaussian components, including a very broad component. A demonstration of all the components used for fitting a particular AGN, namely: SDSS J030417.76+002826.4, is shown in Figure \ref{fitting_demo}. 
 
Since this work concerns the parameters related to the emission line shapes, precise profile measurements were required. We rejected the sources for which emission line detection was not possible significantly. Another issue we encountered was that either the H$\beta$ or the O[III] flux values reported zero values. In this case, the flux ratios become either 0 or infinite, which leads to the rejection of the respective source. It may be recalled that our parent sample of NLSy1 was chosen from \citet{Ojha2020} based upon their X-ray detection where the strength of either the narrow or broad emission lines were not considered in the selection procedure. As a result, many sources with weak emission lines might have been included in the parent sample, which were rejected due to the criteria mentioned above. Finally, out of 206 NLSy1 galaxies, we could fit and get proper measurements of physical quantities for 144 objects, while out of 153 BLSy1  galaxies, we obtained proper measurements for 117 objects.

\subsection{Emission line parameters}
From the multiple Gaussians used in the emission line fit, first, we estimated the parameters characterizing the $H\beta$ emission line. In the case of NLSy1 galaxies, we picked up the single broad Gaussian component for the FWHM of $H\beta$, while we took the average of the broad Gaussian components in the case of BLSy1  galaxies. The broad Gaussian component represents the emission coming from the BLR, while the narrow component is representative of the Narrow Line Region (NLR) emission. The flux of the $H\beta$ emission line was calculated by integrating the flux between 4700\AA{} and 4920\AA{}. We calculated the equivalent width (EW) of the emission line using the same wavelength window, and the monochromatic luminosity at 5100\AA{} ($L_{5100}$) was obtained from the fit.  The iron strength ($R_{fe}$), a crucial parameter responsible for driving the variations in the properties of AGN, was calculated as the ratio of area covered by the broad Fe line between 4433\AA{} and 4684\AA{} and the flux of the $H\beta$ emission line. Further, to understand the influence of NLR emission in driving the variations in the properties of both the types of galaxies, we estimated the R5007 parameter, which is the flux ratio of the narrow [OIII] component at 5007\AA{} and the broad $H\beta$ emission line \citep{Gaur2019}. Moreover, the ratio of broad components of $H\alpha$ and $H\beta$  was calculated using the flux of the broad components of the two emission lines. The Gaussian components used here were similar to those used for estimating the FWHM of the $H\beta$ emission line. We estimated the uncertainties in the respective parameters by fitting the individual spectra in 100 iterations using a Markov Chain Monte Carlo (MCMC) implementation in order to build the distribution of fitted parameters. We then took the 16th and the 84th percentile values as the uncertainty range in these parameters. The typical uncertainties in the FWMH were 10\%, while in the EW, flux, and others, it was in the range of 10\% to 30\%. To get the uncertainties in the flux ratios, we propagated the uncertainties in the flux of the respective emission lines.

\subsection{Asymmetry in the emission line}
 \label{ai_sec}
 To characterize the line shapes in the $H\beta$ emission-line profiles, we estimated the asymmetry index (AI) and Kurtosis Index (KI) for all the AGN  in our sample. AI has been calculated with different flux values in the recent past \citep{Marziani1996, Brotherton1996, Du2018}. We followed the technique used in the previous works \citep[see][]{Heckman1981, Marziani1996, Wolf2020} and chose a combination of 75\% and 25\% flux values in order to estimate AI. The wavelengths at which the emission line profile constructed by adding the broad Gaussian components reach the 75\% and 25\% of the flux values are recorded, and AI, as well as KI, are calculated a
 
\begin{equation}
\centering
{\bf
\begin{array}{l}
 {\displaystyle AI=\frac{\lambda_R^L + \lambda_B^L-2\lambda_0}{ \lambda_R^L - \lambda_B^L}}\\\\\\
 {\displaystyle KI=\frac{\lambda_R^H - \lambda_B^H}{\lambda_R^L - \lambda_B^L}}
\end{array}}
\label{ai_eqn}
\end{equation} 

where $\lambda_B^H$ and  $\lambda_B^L$ are the wavelength values at which the blue wing of the flux reaches 75\% and 25\% of the peak flux, respectively; while  $\lambda_R^H$ and $\lambda_R^L$ the wavelength values at which the red wing of the emission profile reaches at 75\% and 25\% of the peak flux, respectively. $\lambda_0$ is the peak wavelength for the H$\beta$ emission line. The emission profile constructed by adding the broad Gaussian components of the emission line was used to estimate these parameters. The value of AI can range from $-$1 to 1, while the value of KI can range from 0 to 1. Negative AI means blue asymmetry, characteristic of the outflow component, while positive AI means red asymmetry, characteristic of the inflow component. Thus, AI can be used as an indicator of gas dynamics in the line emitting regions of AGN. We estimated the uncertainties in the AI and KI values by propagating the uncertainties in FWHM. The resultant typical uncertainties in AI and KI values were found out to be 0.05

\subsection{SMBH mass, Eddington ratio and X-ray photon indices}

The SMBH mass has been estimated using various empirical relations in the recent past \citep[see][for a review]{Shen2013}. The SMBH mass has been measured using stellar dynamics for nearby quiescent galaxies. The single epoch SMBH mass estimation technique is based on the scaling relations obtained from local galaxy stellar velocity dispersion \citep{Gebhardt2000}. Reverberation mapping \cite[see][]{Bahcall1972, Blandford1982} based SMBH masses provide tighter constraints, and thus far, this technique has been the only reliable SMBH mass estimation method up to higher redshifts \cite[][]{Bentz2009}. This technique assumes the virial motion of BLR clouds around the central SMBH and the Radius-Luminosity (R-L) relation holding true for the entire type-1 AGN population. The equation is:\\

\begin{equation}
 \log\left[\frac{M_{BH}}{M\odot}\right]=0.91+0.5\log \left[\frac{\lambda L_{\lambda}}{10^{44} erg s^{-1}}\right]+ 2\log\left[\frac{FWHM}{km s^{-1}}\right] 
\end{equation}

\vspace{0.5cm}
This scaling equation has been obtained from \citet{Peterson2006}, and 0.91 and 0.5 are the scaling constants to be used in the case of H$\beta$ emission line, the FWHM is in the units of km s$^{-1}$ and $\lambda L_{\lambda}$ is the luminosity at 5100 \AA{} (L$_{5100}$). This equation is based on the assumption that the radius-luminosity (R-L) relation available for a set of approximately 120 reverberation mapped AGN so far \citep{Bentz2009, Yu2020} holds for the type 1 AGN in general. We estimated the SMBH mass using the broad component of the FWHM of the $H\beta$ emission line for all the AGN. The NLSy1 galaxies had smaller SMBH masses, owing to the small FWHM of the  $H\beta$ emission line. The uncertainties in SMBH mass were estimated by propagating the errors in FWHM and the Luminosity values. We obtained a typical uncertainty of around 0.3 to 0.4 dex, which is consistent with the values obtained in the past works \citep{Peterson2006, Shen2013}.

We estimated the bolometric luminosity ($L_{bol}$) for the AGN using the empirical relations available in the literature. The scaling for the bolometric luminosity is 9.1 times the one estimated at 5100\AA{} \citep[see][]{kaspi2000}. Using $L_{bol}$ and the obtained SMBH masses, we estimated the Eddington ratio $R_{\textsc{edd}}$. The Eddington ratio is defined as the ratio of the bolometric to the Eddington luminosity ($L_{bol}/L_{\textsc{edd}}$). The equations for  $L_{bol}$ and $L_{\textsc{edd}}$ are as follows:
\begin{equation}
    L_{bol}= 9.1 \times L_{5100};\\\\
    L_{\textsc{edd}}= 1.45\times 10^{38}\times \frac{M_{BH}}{M\odot}  erg/s
\end{equation}

 In \citet{Ojha2020}, $R_{\textsc{edd}}$  was calculated using X-ray observations which were consistent with the estimation obtained from optical parameters; hence we did not attempt to estimate $R_{\textsc{edd}}$  using other methods and used the value obtained from optical spectra only.  The soft X-ray photon indices ($\Gamma_{X}$) were obtained from \citep{Ojha2020}. It has been reported that the NLSy1 galaxies have steeper X-ray spectra, thus a high value of $\Gamma_{X}$ \citep{Waddell2020}. Being one of the fundamental components in the 4DE1 formalism of \citet{Boroson1992}, the comparison of $\Gamma_{X}$ with the physical properties of both the types of galaxies can provide clues to the peculiar behavior of NLSy1 galaxies as we have done in this work. All the estimated parameters, along with the uncertainties, are presented in the form of a table (see Table \ref{table1}).

 \begin{figure*}

\includegraphics[width=8cm,height=6cm]{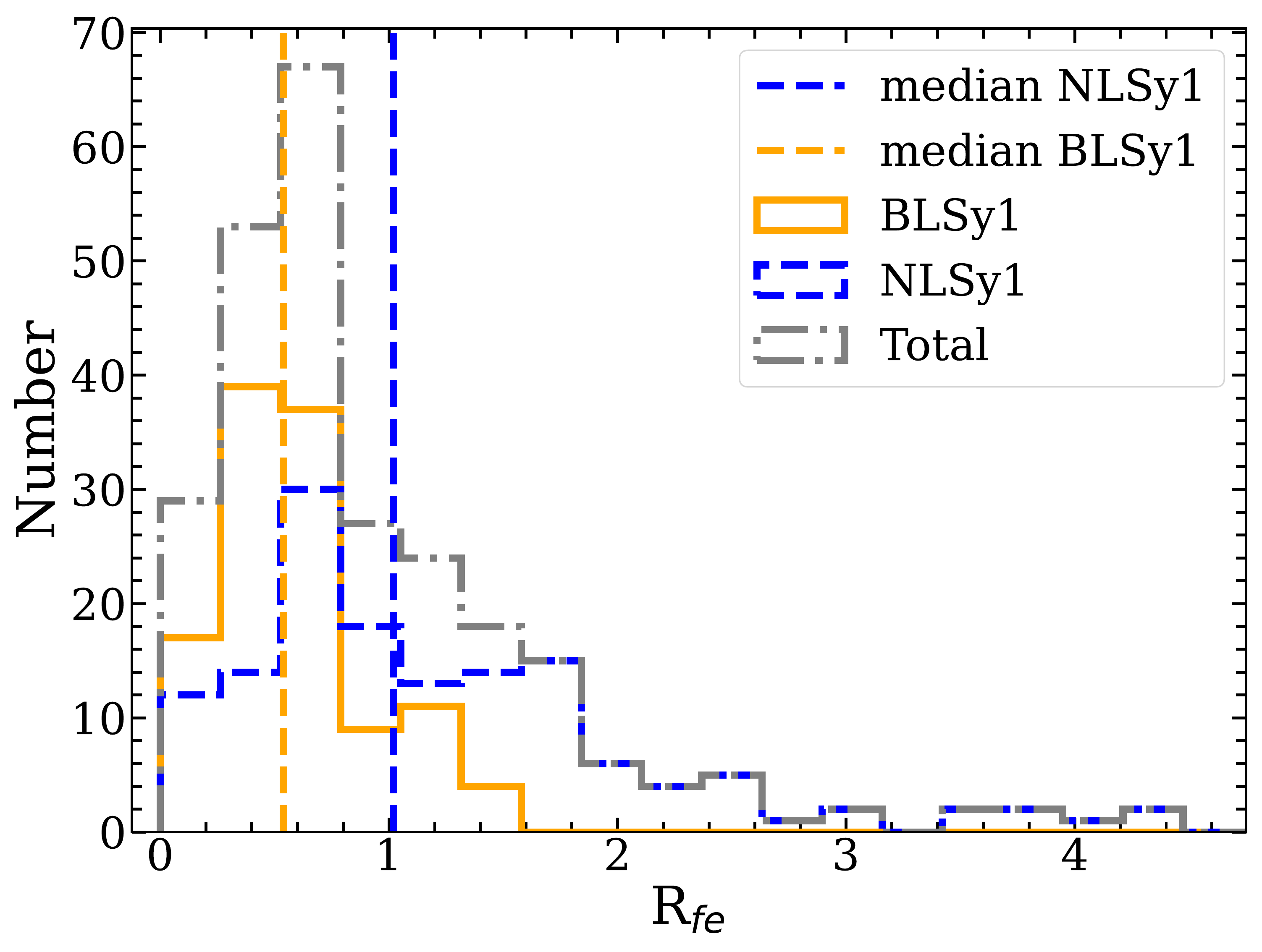}
\includegraphics[width=8cm,height=6cm]{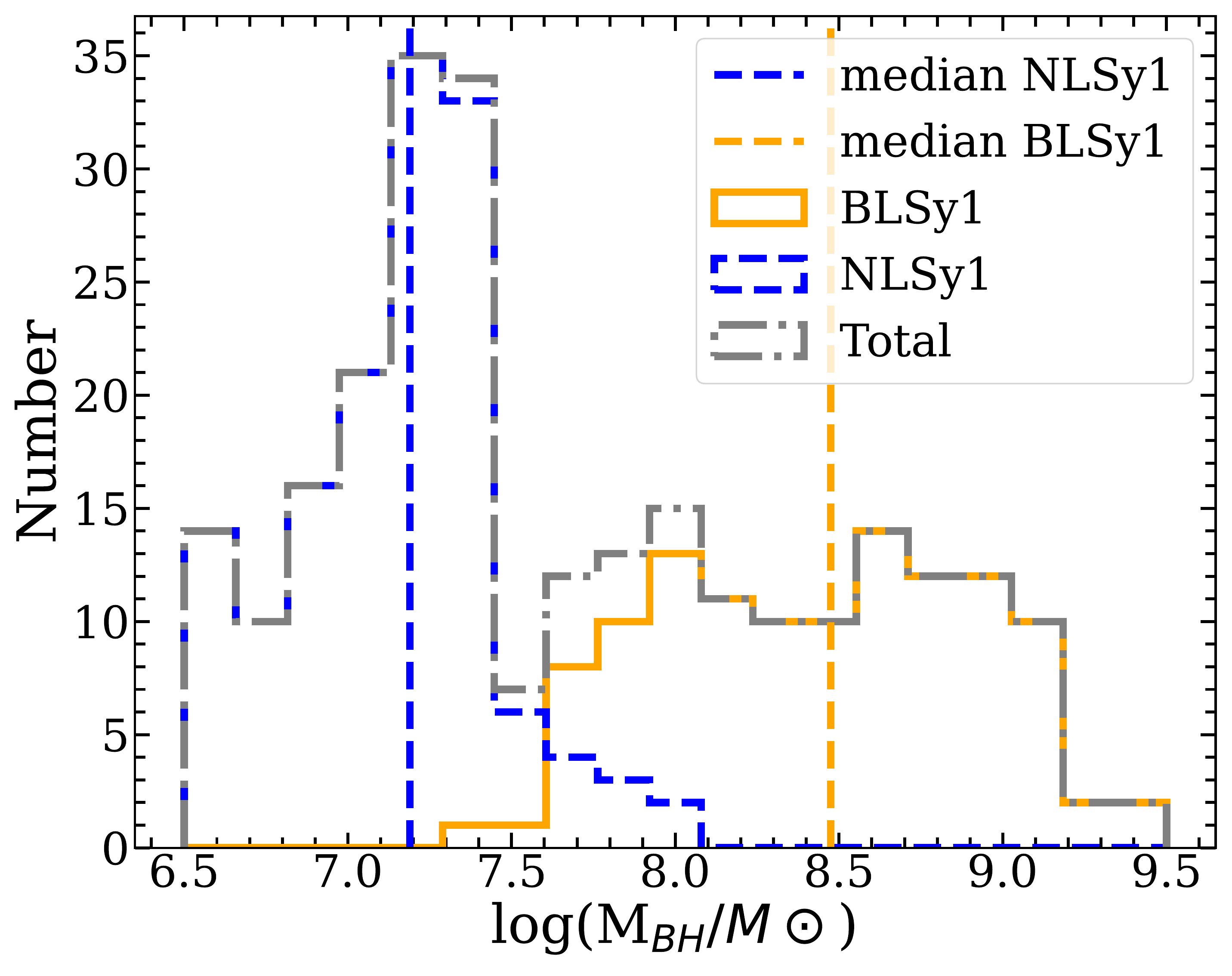}
\includegraphics[width=8cm,height=6cm]{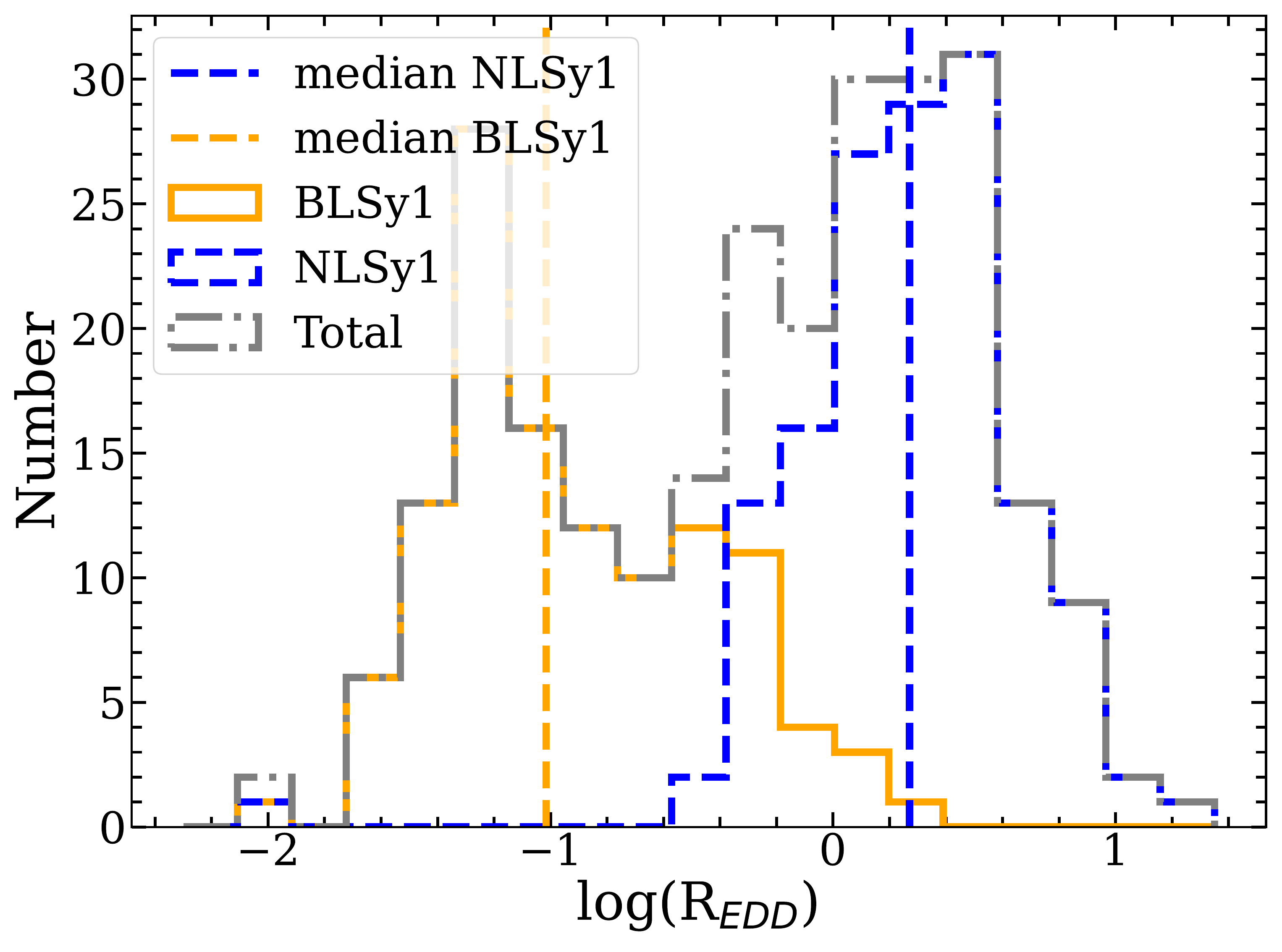}
\includegraphics[width=8cm,height=6cm]{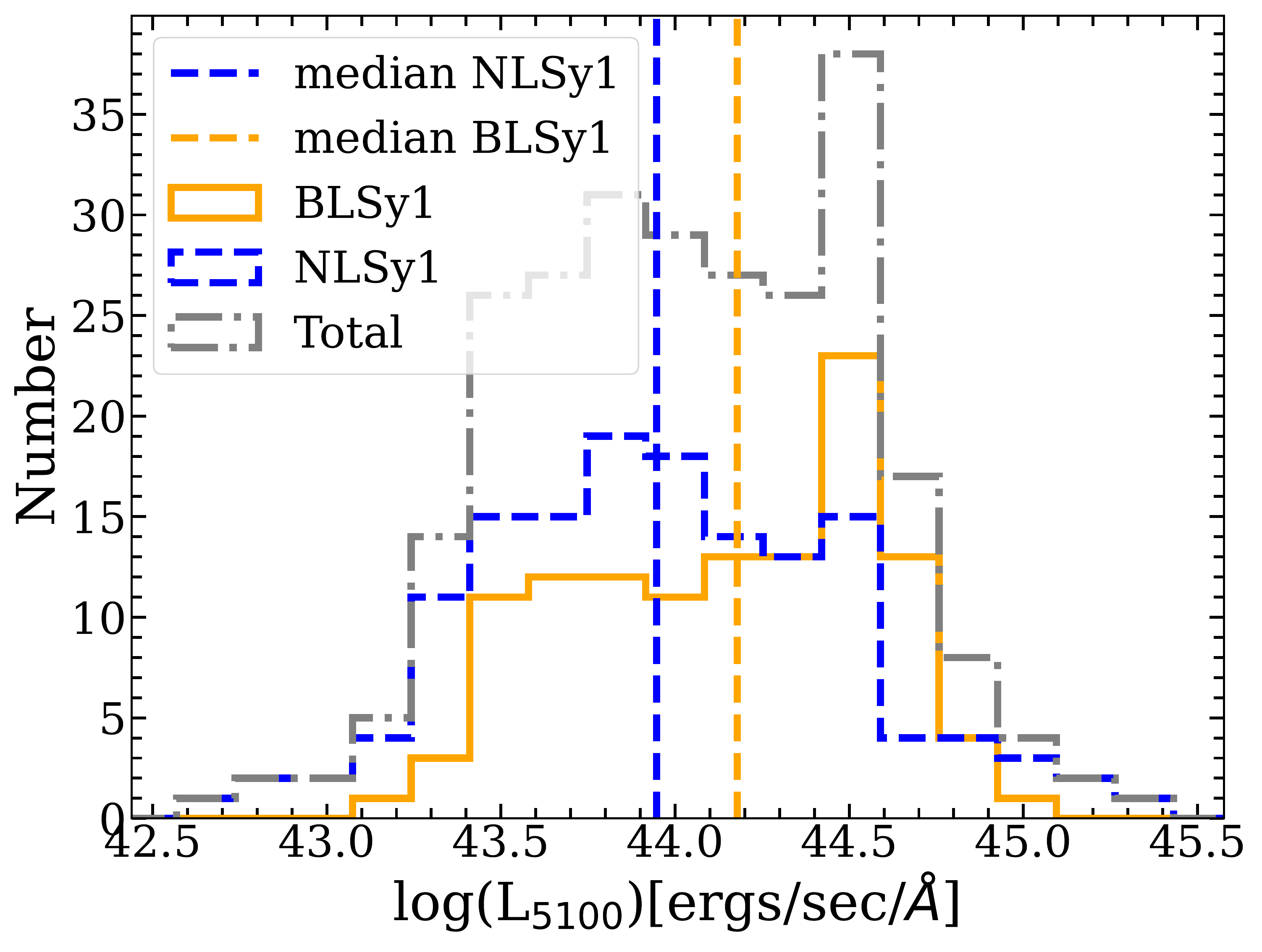}
\includegraphics[width=8cm,height=6cm]{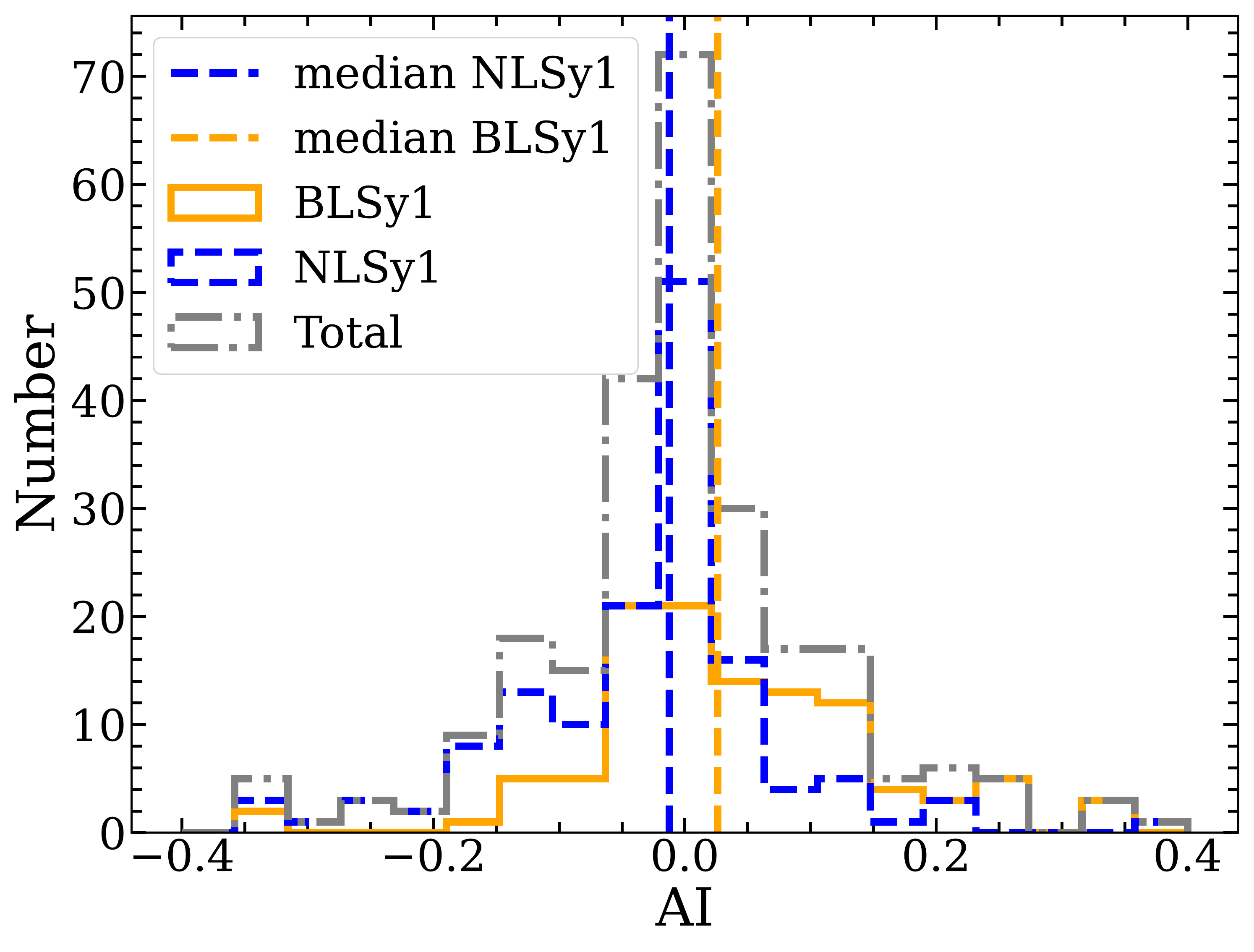}
\includegraphics[width=8cm,height=6cm]{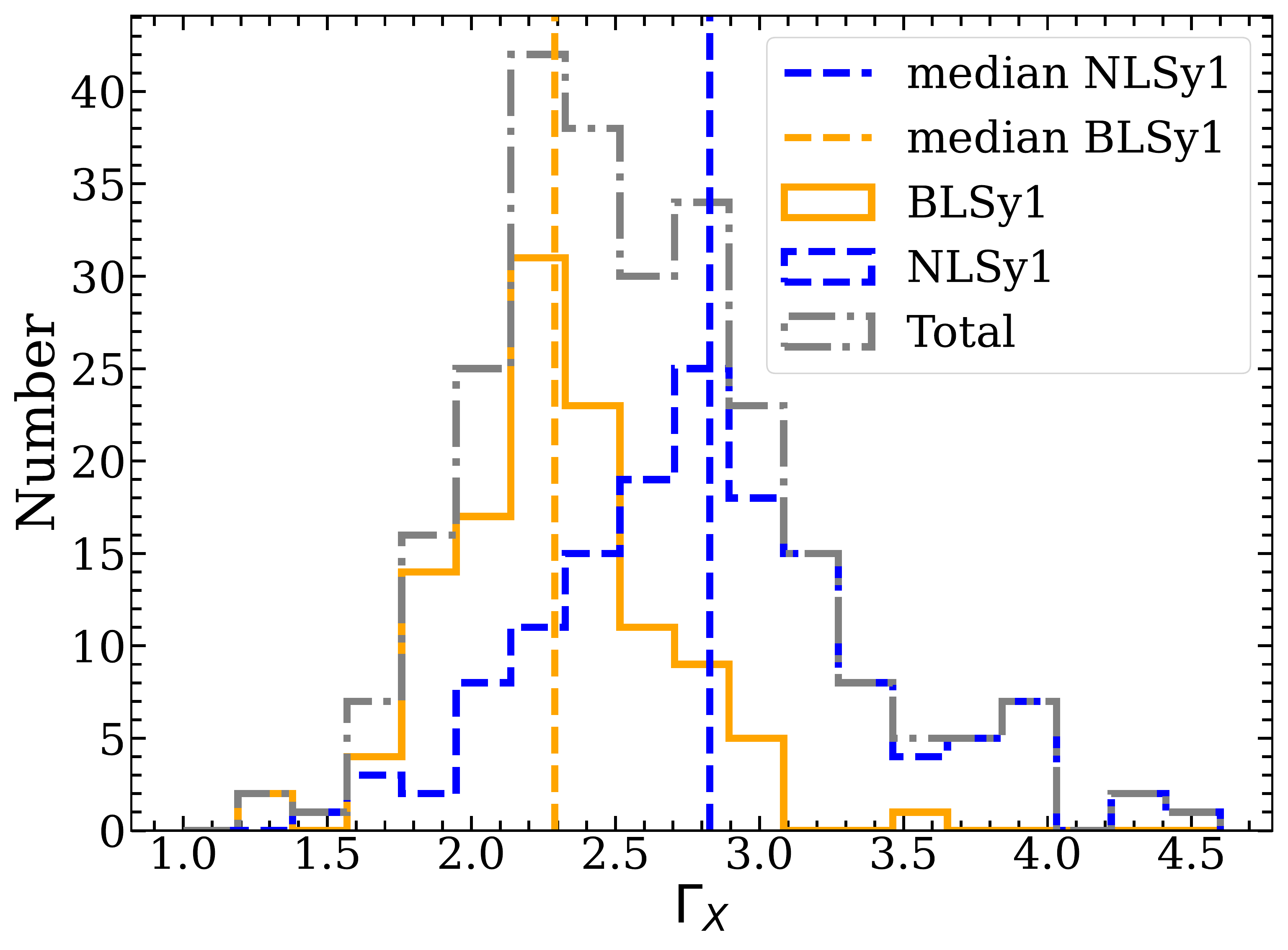}

\caption{Distribution of various physical parameters for the NLSy1 galaxies (blue dashed line), BLSy1  galaxies (solid orange line), and combined sample (dashed grey line). The top-left and right panels show the distribution for iron strength ($R_{fe}$), and the distribution for SMBH mass, respectively, the middle-left and middle-right panels show the distribution for Eddington ratio ($R_{\textsc{edd}}$), and the distribution for luminosity at 5100 \AA{}, respectively. The bottom-left and bottom-right panels show the distribution of AI calculated using Equation \ref{ai_eqn} and the distribution of soft X-ray photon indices ($\Gamma_{X}$), respectively. The median values of the parameters for both types of galaxies are shown as vertical lines.}
\label{param_dist}
\end{figure*}

\begin{figure}
\includegraphics[width=9cm,height=7cm]{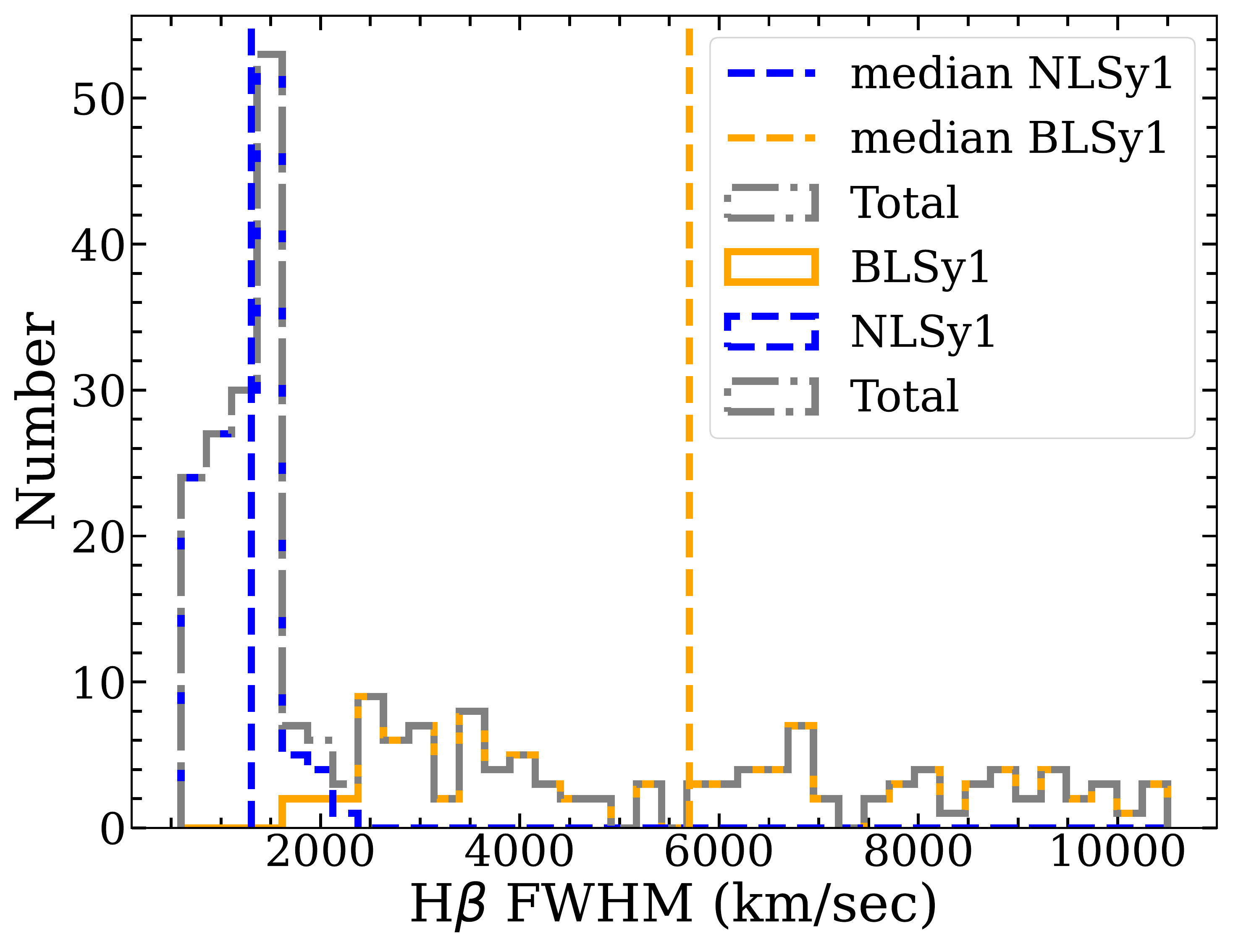}

\caption{The distribution for FWHM of the H$\beta$ emission line for the NLSy1 galaxies (blue dashed line), BLSy1 galaxies (solid orange line), and combined sample(solid grey line). The median values for the BLSy1 and NLSy1 galaxies are shown as blue dashed and orange solid vertical lines, respectively.}
\label{fwhm_dist}
\end{figure}

 \begin{figure*}
 \subfigure[]{\includegraphics[width=8.5cm,height=7cm]{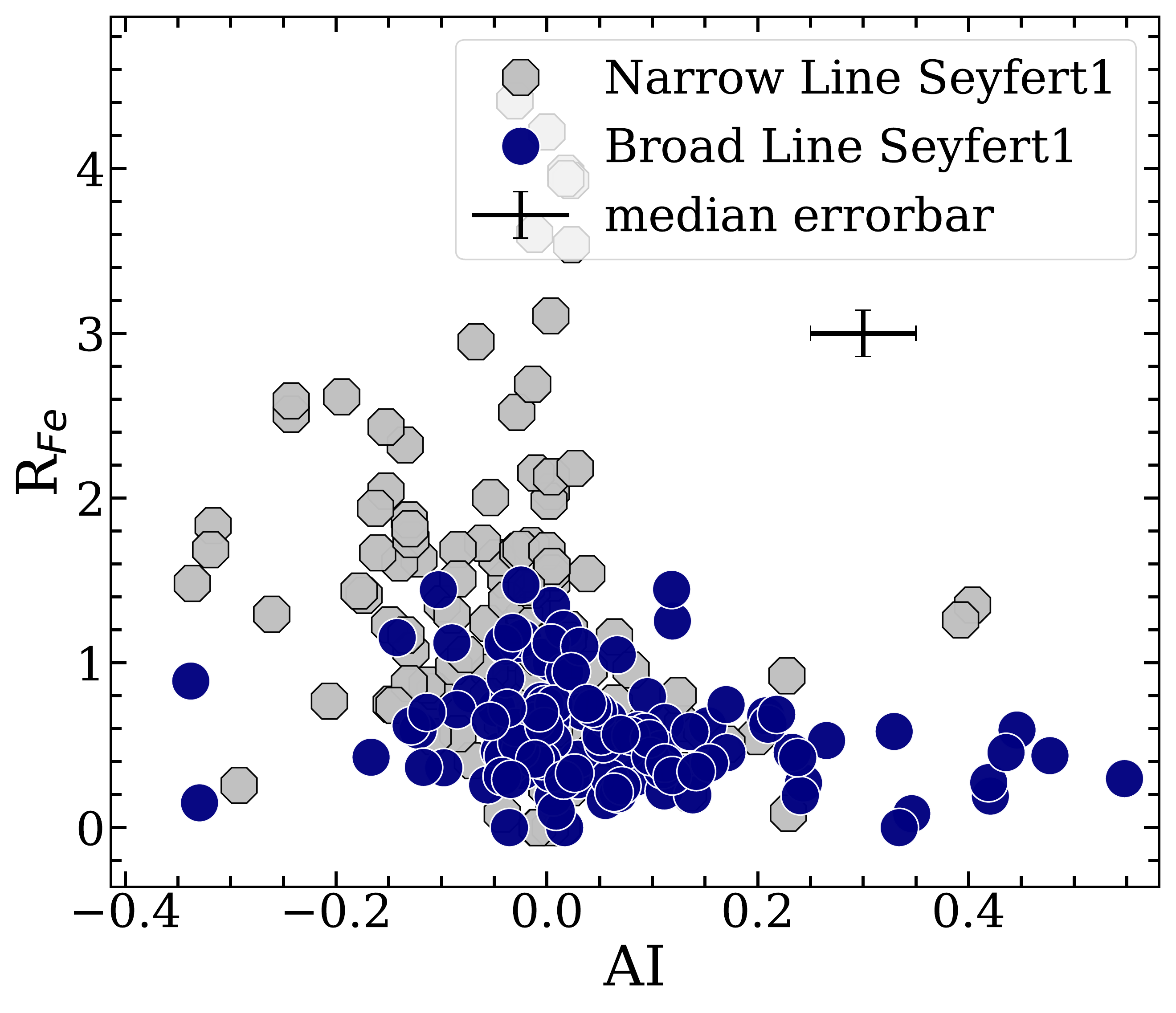}}
 \subfigure[]{\includegraphics[width=8.5cm,height=7cm]{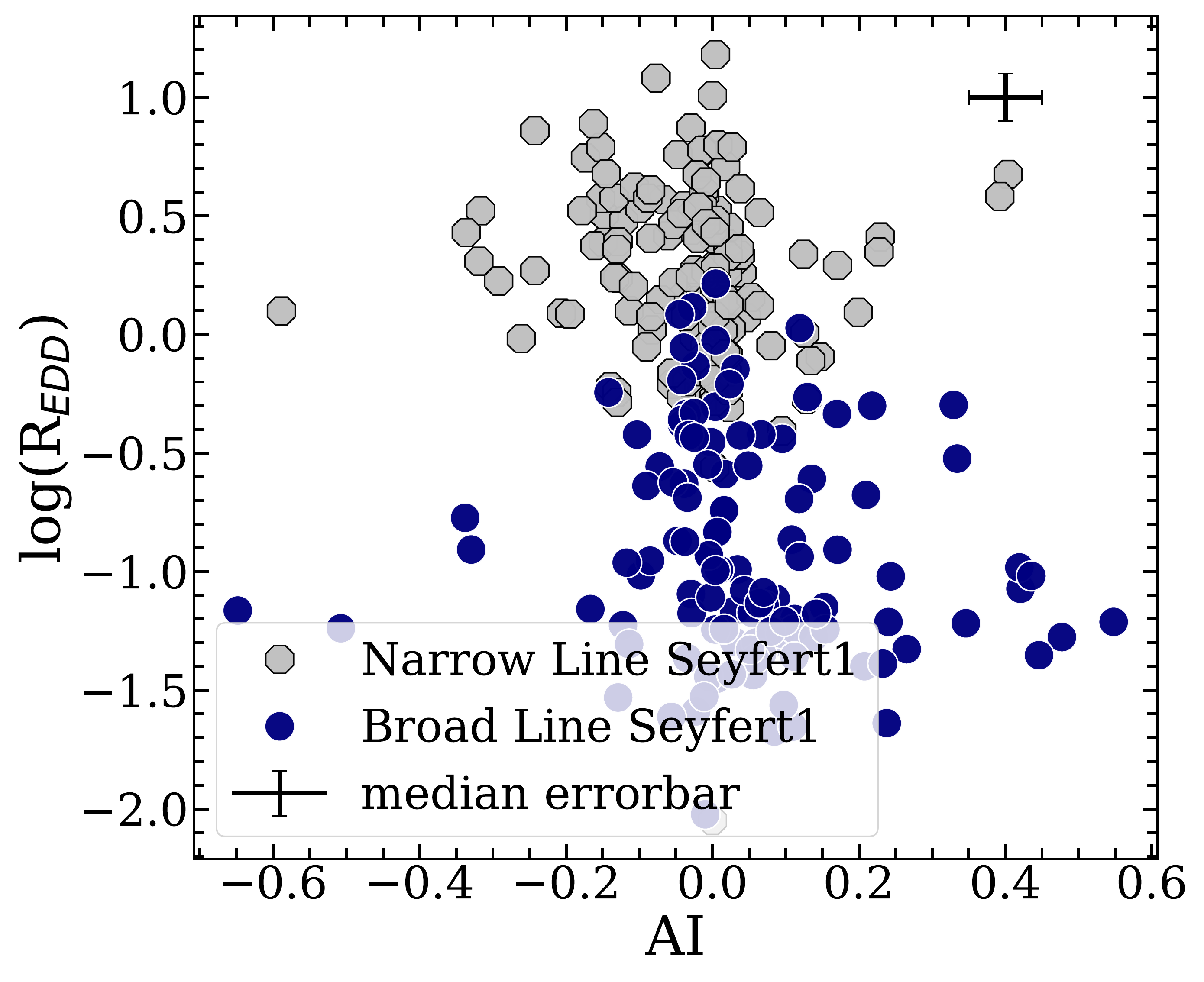}}
 \subfigure[]{\includegraphics[width=8.5cm,height=7cm]{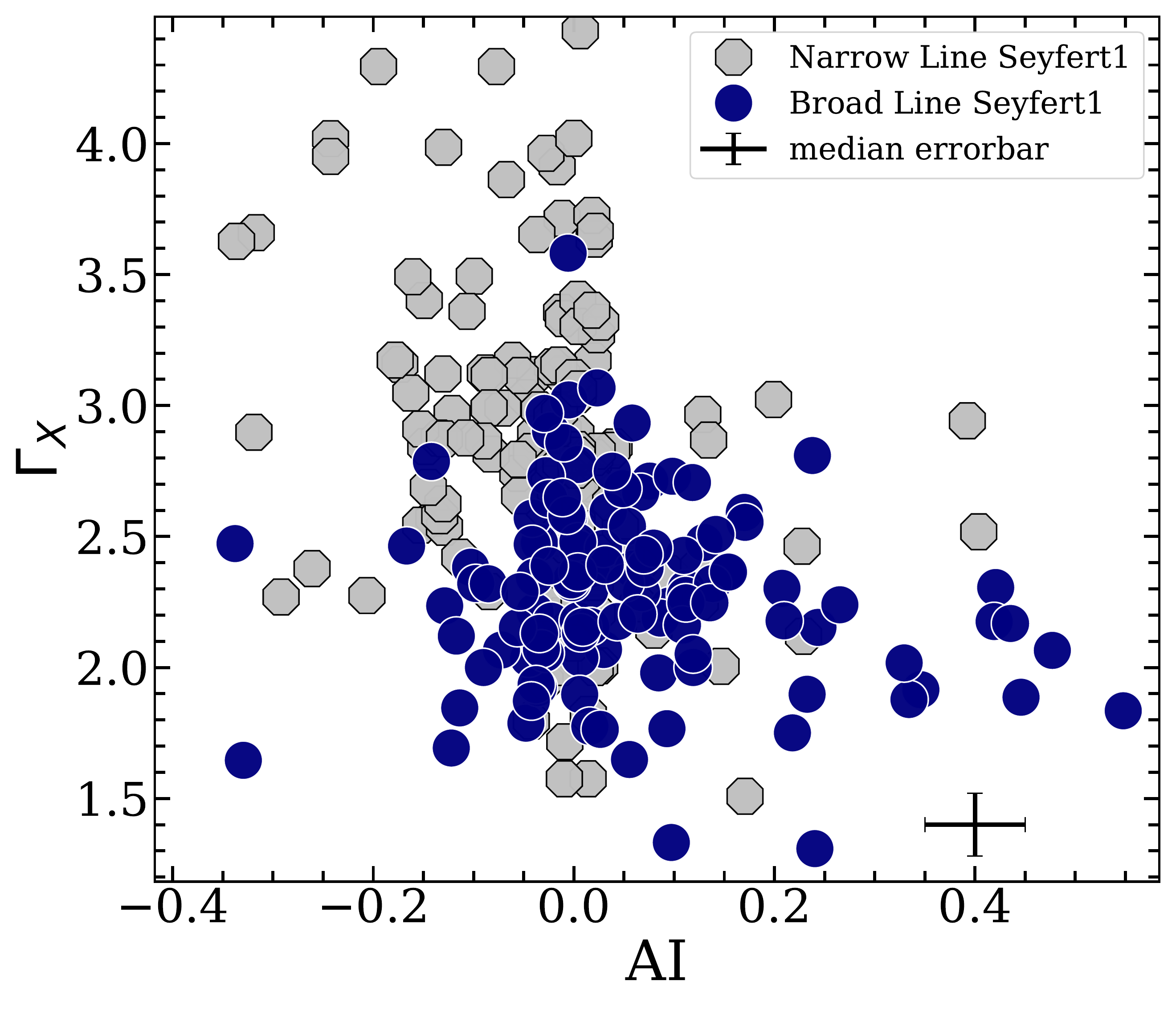}}
  \subfigure[]{\includegraphics[width=8.5cm,height=7cm]{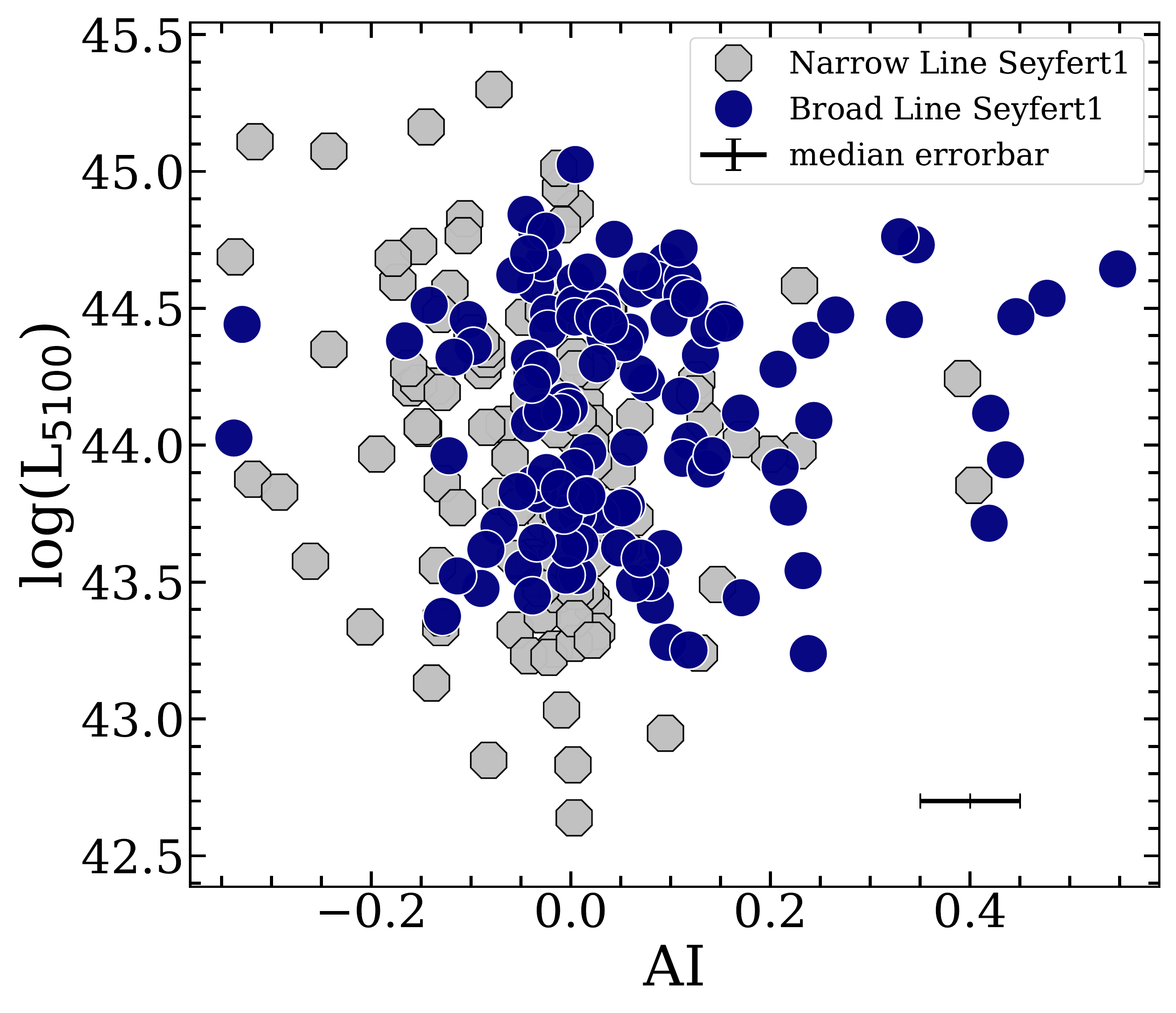}}

 \caption{Correlations between AI and 4 physical parameters for the NLSy1 galaxies (grey circles) and BLSy1  galaxies (blue circles). Relation of AI with $R_{fe}$ is shown in panel (a), with  $log(R_{\textsc{edd}}$)  is shown in panel (b), with $\Gamma_{X}$ in panel (c), and with $L_{5100}$ in panel (d);  representative of the physical processes in the central regions of AGN. The median errorbars on each parameter are also shown.}
\label{figure}
\end{figure*}

\section{Results}
\label{results}

\subsection{Intrinsic distribution of parameters}

The properties of the line emitting regions, i.e., the BLR and NLR of AGN, can be characterized using a set of parameters derived from the optical spectra, while the X-ray photon index provides clues to the innermost regions of the accretion disk. We used 11 observational and physical parameters, namely the FWHM of the $H\beta$ emission line, Equivalent width of $H\beta$ (EW) emission line, the $[OIII]/H\beta$ flux ratio, the $H\alpha/H\beta$ flux ratio, $R_{fe}$, $L_{5100}$, SMBH mass, $R_{\textsc{edd}}$, AI, KI and $\Gamma_{X}$ to understand the diversity in the properties of BLSy1 and NLSy1 galaxies. The distribution for 6 of these parameters is shown in Figure \ref{param_dist}, while the distribution for the FWHM of H$\beta$ emission line is shown in Figure \ref{fwhm_dist}. Orange and blue colors denote the BLSy1 galaxies and the NLS1 galaxies, and the combined sample is denoted by grey color in the histograms. The vertical lines in the middle denote the median value of the respective populations. We found the iron strength in NLSy1 galaxies to be higher than the BLSy1  galaxies for most of the AGN, which is consistent with the past works \citep[see][]{Gaskell2000}. Results based on {\sc cloudy} based simulations also indicate similar phenomenon \citep{panda_rfe}. In the current work, the median  $R_{fe}$ was 1.02 for the NLSy1 galaxies while it was nearly half of that value, 0.53 for the BLSy1  galaxies. Eddington ratio is one of the important classifiers for NLSy1 galaxies. We found out that the Eddington ratio is higher for the NLSy1 galaxies, the median value of log($R_{\textsc{edd}}$) being 0.22 while the median log($R_{\textsc{edd}}$) was $-$1.02  in the case of BLSy1  galaxies. Furthermore, we found out that the SMBH masses are lower for the NlSy galaxies with the median value of log($\frac{M_{BH}}{M\odot}$)= 7.18, while in the case of BLSy1  galaxies, the median log($\frac{M_{BH}}{M\odot}$) was higher at 8.47. 

Based on the uncertainties observed in AI in Sect. \ref{ai_sec}, we set a criterion of AI $\leq -$0.05 and AI $\geq$ 0.05 as significant blue and red asymmetries, respectively. We found out that more NLSy1 galaxies show blue asymmetries as compared to the BLSy1 galaxies. In the current sample, 46 NLSy1 galaxies show significant blue asymmetric profiles compared to 17 BLSy1 galaxies, while 20 NLSy1 galaxies show red asymmetric profiles compared to 52 BLSy1 galaxies. Blue asymmetries indicate that there is outflowing gas arising from that region. In recent works, the sources with high  $R_{fe}$ values have been known to show blueward asymmetries \citep[see][]{Ganchi2019, Wolf2020}. The exact relation between the two is not clear, but high accretion rates could likely play a part as low accretors typically possess red asymmetries \citep{Zamfir2010}.  

The broad $H\alpha$/$H\beta$ flux ratio has been used in the literature to understand the influence of the dust on the broad emission lines \citep{Dong2008}. Our sample shows slightly different behavior in both types of galaxies, with the median value for NLSy1 galaxies being 2.01, while the median value for BLSy1  galaxies is a bit higher at 2.47.

We performed a two-sample Kolmogorov Smirnov (KS) test \citep{Massey1951} on the distributions of the parameters in order to check the similarity of both populations. If the p-value is small, then it implies that the two populations are significantly different. The p-value was $\leq 0.01$ for all but two of the parameters in the sample. This signifies that the distributions were not drawn from a common sample (see Table \ref{ks_test_results}). Only $L_{5100}$ and the $[OIII]/H\beta$ flux ratio had a p-value greater than 0.01, which rejects the null hypothesis that the sources are drawn from the same parent sample. As our sources were in a similar luminosity range, we did not expect the luminosity characterized by $L_{5100}$ to be statistically different in the two populations. However, in this sample, the $[OIII]/H\beta$ flux ratio may not be a parameter responsible for the diversity between the two types of galaxies, as is evident from the very similar median values for both the types of galaxies as well.

\subsection{Spearman rank correlations}
\begin{figure*}
\includegraphics[width=11cm,height=7cm]{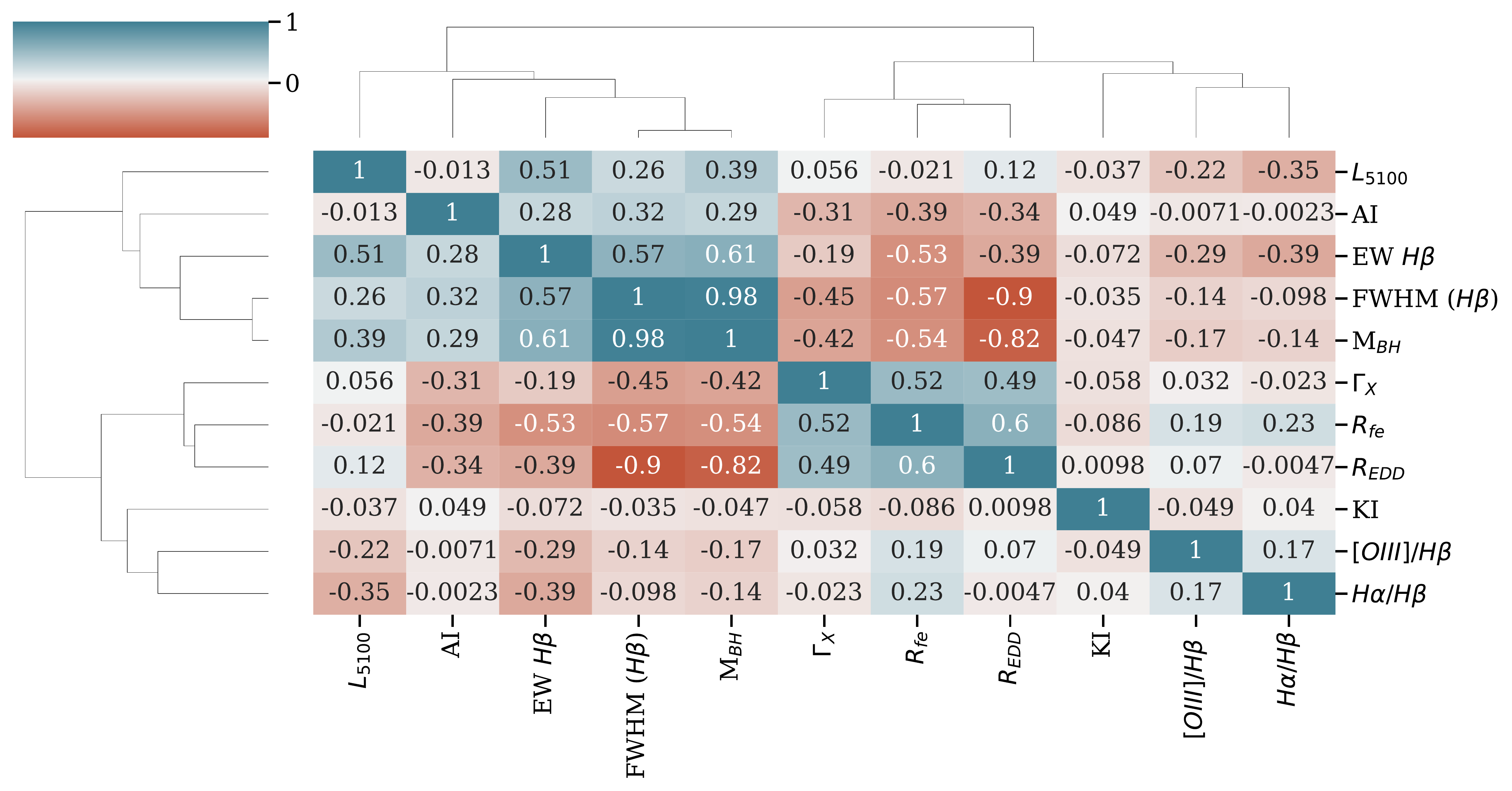}

\caption{The Spearman rank correlation matrix along with a cluster map for the entire sample of 144 NLSy1 galaxies and 117 BLSy1  galaxies. A combination of 11 parameters namely the $H\beta$ FWHM (broad component), Equivalent width of $H\beta$ (EW), the $[OIII]/H\beta$ flux ratio, the $H\alpha/H\beta$ flux ratio, iron strength ($R_{fe}$), $L_{5100}$, SMBH mass, $R_{\textsc{edd}}$ , asymmetry index (AI), kurtosis index (KI), and Soft X-ray photon index ($\Gamma_{X}$) have been used in this analysis. The cluster map denotes the linking of various parameters with each other. The orange rectangles denote the anti-correlations, while the blue rectangles denote the positive correlations. The Spearman rank correlation coefficient is written on the rectangular boxes.}
\label{cluster_all}
\end{figure*}

\begin{figure*}
 \hspace{-1cm}
 \subfigure[Spearman correlations for NLSy1 galaxies. ]{\includegraphics[width=9cm,height=6cm]{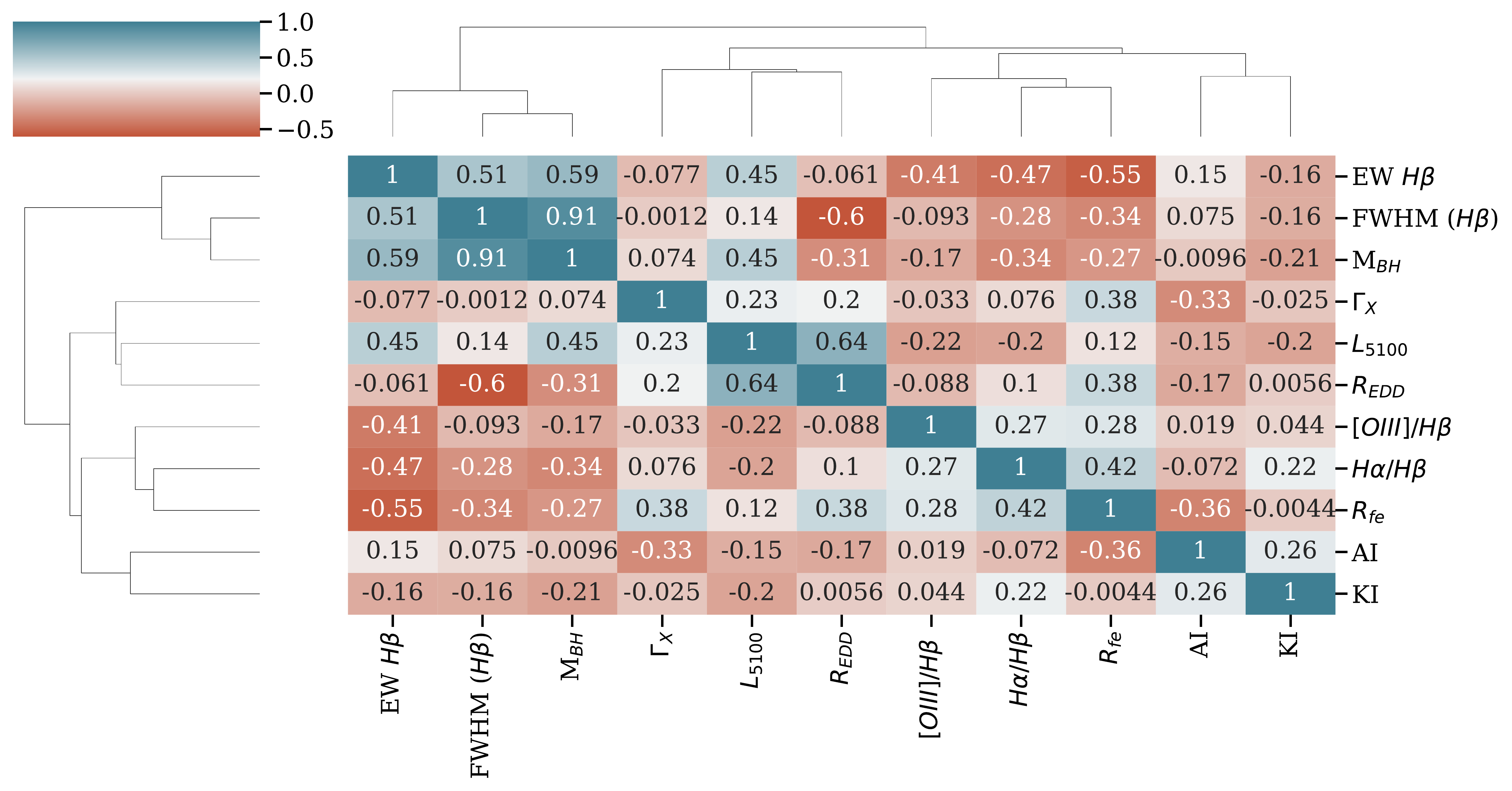}}
    \subfigure[Spearman correlations for BLSy1  galaxies.]{\includegraphics[width=9cm,height=6cm]{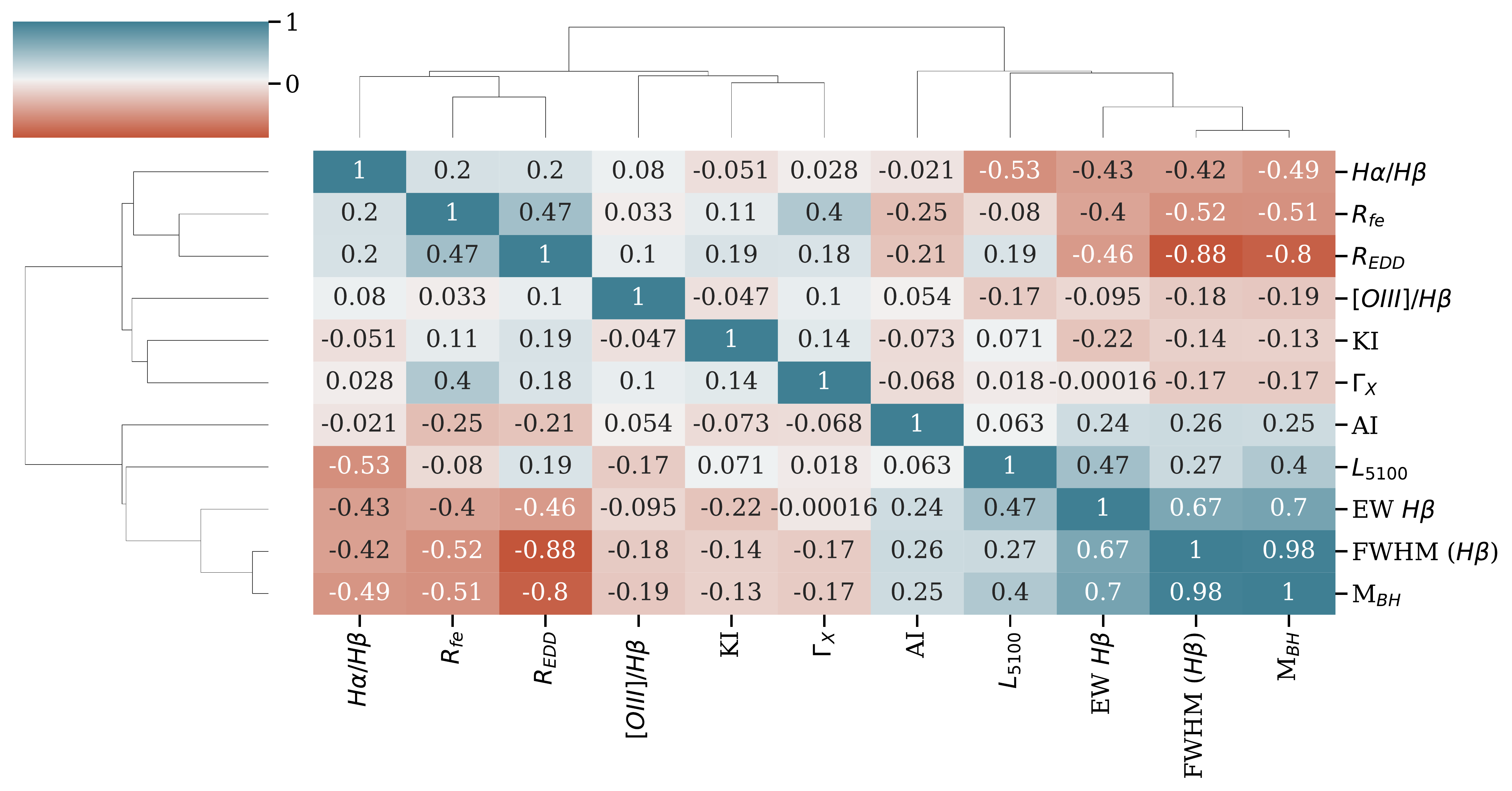}}
    \caption {The  Spearmann rank correlation matrix along with a clustermap presented individually for the NLSy1 galaxies (left) and BLSy1  galaxies (right).  The correlations and anti-correlations are denoted in the same way as Figure \ref{cluster_all}.}

     \label{cluster_both}
 \end{figure*}

We calculated the Spearman rank correlation coefficients among all the parameters obtained for the entire sample. We performed the correlation analysis on NLSy1 and BLSy1  galaxies individually and on the combined population. The results are presented in Figures \ref{cluster_all} and \ref{cluster_both}. We also included a cluster map that separates the parameters into two distinct groups, one dominated by FWHM, other by $R_{fe}$ in all three cases. The FWHM of $H\beta$ correlates negatively with $R_{fe}$ for both the classes, with the NLSy1 galaxies having a correlation coefficient of $-$0.34 while the BLSy1  galaxies have a correlation coefficient of $-$0.52. When the cross-correlation is run on the entire sample, the anti-correlation coefficient is $-$0.57. This is a well-known relation and forms the backbone of the quasar main sequence \cite[see][and references therein]{Marziani2018}. $R_{\textsc{edd}}$ is the most distinguishing parameter in our analysis. This parameter separates the population of NLSy1 galaxies and BLSy1  galaxies into two distinct classes, which is also apparent from the distribution in Figure \ref{param_dist}. While  $R_{\textsc{edd}}$  anti-correlates weakly with the SMBH mass (correlation coefficient of $-$0.30) in the case of NLSy1  galaxies, a strong anti-correlation is seen between the two in the case of BLSy1 galaxies, with a correlation coefficient of $-$0.8. However, between the FWHM and $R_{\textsc{edd}}$, we observe a similar anti-correlation for both types of galaxies, with the correlation coefficient being $-$0.6 for the NLSy1 galaxies and $-$0.8 for the BLSy1 galaxies.  For the entire sample, a strong anti-correlation between $R_{\textsc{edd}}$ and $H\beta$ FWHM is seen, with the anti-correlation coefficient of $-$0.9. Moreover, we also find a correlation between $R_{\textsc{edd}}$  and $R_{fe}$ in the case of NLSy1 galaxies (correlation coefficient of 0.38) as well as the BLSy1  galaxies (correlation coefficient of 0.47).

The SMBH mass significantly correlates with the $H\beta$ FWHM in all the cases. Since the SMBH mass is a derived quantity based on the $H\beta$ FWHM among other parameters, it is expected to correlate to some degree. The Balmer decrement calculated using the ratio of the area of the $H\alpha$ and $H\beta$ emission lines, anti-correlates with the $H\beta$ FWHM in both the cases with the correlation coefficient of $-$0.28 in NLSy1 galaxies and a correlation coefficient of $-$0.42 in the case of BLSy1 galaxies. In \citet{Dong2008}, it was observed that the ratio does not correlate much with the AGN physical properties. In their analysis for a sample of 446 low redshift AGN, the $H\alpha/H\beta$ flux ratio shows a very weak correlation of 0.078 with $H\beta$ FWHM. Our results are in deviation of this result.
 
$\Gamma_{X}$  is anti-correlated with the FWHM of H$\beta$ emission line if we take the entire sample into account, which is consistent with the results obtained from the sample analyzed by \citet{Grupe2004}. However, there is no significant correlation between the two if it is calculated separately. $\Gamma_{X}$ also correlates with $R_{\textsc{edd}}$ and  $R_{fe}$, which indicates that the inner regions of the accretion disk have a role in the diversity of the parameters in the two types of galaxies.

The FWHM of H$\beta$ correlates positively, although weakly, with AI. The correlation coefficient is 0.26 in the case of BLSy1  galaxies, while it is a meager 0.07 in the case of NLSy1 galaxies. Also, it anti-correlates with $R_{fe}$ in both the cases of NLSy1 galaxies and the BLSy1  galaxies. AI shows weak positive correlations with the SMBH mass in both cases, mimicking the behavior of H$\beta$  FWHM. Correlation between the $R_{\textsc{edd}}$ and blueshift in the C {\sc iv} emission line has been observed in \citet{Sulentic2017}. However, in our study, we have found that AI is anti-correlated with the $R_{\textsc{edd}}$ in the case of NLSy1 galaxies, while it is weakly correlated in the case of BLSy1 galaxies. There has been a postulation to use AI in the emission profile as a surrogate parameter in the 4DE1 formalism \citep{Zamfir2010}. The correlations between AI and other parameters mimic the behavior of FWHM, yet the correlation between these two parameters itself is not very strong. Thus, based on these results, we cannot conclusively say that AI can be used as a surrogate parameter in the 4DE1 formalism.

\begin{table*}

\begin{tabular}{lrrc}
\hline
{\bf Parameter}   &  \multicolumn{2}{c}{\bf Median value} &     {\bf p-null}\\
\hline
\\ & {\bf NLSy1}  & {\bf BLSy1}  & \\
\hline

$H\beta$ FWHM (km s$^{-1}$)                                              & 1304.4 & 5699.8 & 1.67$\times$10$^{-98}$ \\
$H\beta$ EW (\AA)               
&42.42 & 72.03& 3.45$\times$10$^{-07}$ \\
$[OIII]/H\beta$                                          
&0.20 & 0.16 & 3.2$\times$10$^{-02}$ \\
$H\alpha/H\beta$                                           
&2.00 &       2.47            & 4.0$\times$10$^{-03}$                   \\
$R_{fe}$                                                   
&   1.02 &      0.53           & 3.94$\times$10$^{-12}$                    \\
log($L_{5100}$)                                                 
&  43.94 &    44.17               & 1.17$\times$10$^{-02}$                         \\
log($M_{BH}/M\odot$)                                             
&   7.18   &    8.47           &4.48$\times$10$^{-15}$                     \\
log($R_{\textsc{edd}}$)                                                  &   0.27    &    $-$1.02          & 6.91$\times$10$^{-58}$                     \\
AI                                                         
&  $-$0.01	&         0.02          & 2.56$\times$10$^{-07}$                     \\ 
KI                                                         
&     0.30    &   0.28        & 3.37$\times$10$^{-04}$                    \\
$\Gamma_{X}$                                           
&    2.83 &       2.29          & 1.99$\times$10$^{-15}$                    \\
\hline
\end{tabular}

\caption{ The median values for all the 11 parameters for NLSy1 galaxies (column 2) and BLSy1  galaxies (column 3) are shown in this table. Column 4 shows the {\it p-null} values as a result of a two-sample Kolmogorov Smirnov (KS) test performed on the sample of NLSy1 galaxies vs BLSy1  galaxies. The low values of the {\it p-null} values signify that the distributions are not part of the same parent population.}
\vspace{0.6cm}
\label{ks_test_results}
\end{table*}

\subsection{Principal component analysis}

Principal Component Analysis (PCA) is a technique used to reduce the dimensionality of a data set. PCA converts these observable parameters into principal components which are orthogonal to each other. These components are known as Eigenvectors, and these Eigenvectors reproduce the entire data set. The first Eigenvector (EV1) accounts for the maximum variance, followed by the subsequent orthogonal Eigenvectors. This technique is useful while looking for key parameters responsible for variability in a large sample like the one used here. PCA has often been used in the statistical study of the quasars \citep[see][]{Boroson1992,Brotherton1996,Boroson2004,Grupe2004,Xu2012,Jarvela2015,Waddell2020, Wolf2020}.

\begin{table*}
\centering

\begin{tabular}{lrrrrrrrrrr}
\hline

{\bf Parameters}                   & \multicolumn{5}{l}{\bf Narrow line Seyfert 1 (NLSy1) galaxies} & \multicolumn{5}{l}{\bf Broad Line Seyfert 1 (BLSy1)  galaxies} \\
                   
\hline
                & EV1       & EV2       & EV3       & EV4       & EV5      & EV1       & EV2       & EV3       & EV4      & EV5      \\
\hline

FWHM $H\beta$                & 0.433  & $-$0.14  & $-$0.1   & $-$0.449 & $-$0.028 &   0.428  & $-$0.272 & $-$0.023 & $-$0.049 & 0.185  \\
EW $H\beta$                  &0.418  & 0.025  & 0.122  & $-$0.026 & $-$0.013 &   0.34   & $-$0.105 & $-$0.364 & 0.002  & 0.17   \\
$[OIII]/H\beta$  &          $-$0.304 & $-$0.026 & $-$0.01  & $-$0.553 & $-$0.048 &  $-$0.25  & $-$0.505 & $-$0.263 & $-$0.149 & $-$0.044 \\
$H\alpha/H\beta$             & $-$0.367 & $-$0.016 & 0.031  & $-$0.428 & 0.111  &   $-$0.315 & $-$0.403 & 0.15   & $-$0.032 & $-$0.143 \\
$R_{fe}$                     &$-$0.376 & 0.152  & 0.1    & $-$0.307 & 0.009  &   $-$0.306 & 0.144  & $-$0.215 & $-$0.005 & 0.509  \\
log($L_{5100}$)                  & 0.185  & 0.248  & 0.621  & $-$0.147 & $-$0.054   & 0.184  & 0.179  & $-$0.7   & $-$0.108 & $-$0.205 \\
$log(M_{BH}/M_{\odot})$          &0.449  & $-$0.045 & 0.102  & $-$0.403 & $-$0.03    & 0.455  & $-$0.194 & $-$0.06  & $-$0.053 & 0.082  \\
log($R_{\textsc{edd}}$)        & $-$0.121 & 0.319  & 0.619  & 0.156  & $-$0.035   & $-$0.361 & 0.323  & $-$0.385 & $-$0.002 & $-$0.212 \\
AI                           & $-$0.087 & $-$0.63  & 0.296  & 0.056  & $-$0.025  & 0.071  & $-$0.204 & $-$0.099 & 0.668  & $-$0.54  \\
KI                           &$-$0.083 & $-$0.627 & 0.305  & 0.052  & $-$0.029   & 0.117  & 0.082  & 0.142  & $-$0.687 & $-$0.516 \\
$\Gamma_X$                & $-$0.062 & 0.015  & $-$0.069 & 0.001  & $-$0.989 &   $-$0.237 & $-$0.504 & $-$0.238 & $-$0.203 & 0.064 \\

\hline

\end{tabular}
\caption{  The projections of the first 5 Eigenvectors on the physical parameters for both types of galaxies obtained using the PCA method.}
\label{Table3}
\end{table*}

We performed the PCA using Python's sklearn\footnote{\url{https://scikit-learn.org/stable/}} package on the 11 calculated observational and physical parameters from the current sample to understand the correlation between these parameters in the Eigenvector space. We performed individual PCA on NLSy1 galaxies and BLSy1  galaxies and then jointly on the entire sample. We have included the line shape parameters, namely AI and KI, in the PCA analysis for the first time, keeping the separation between the NLSy1 and BLSy1 galaxies in mind.
 
\begin{figure*}
\hspace{-1cm}
\subfigure{\includegraphics[width=14cm,height=8cm]{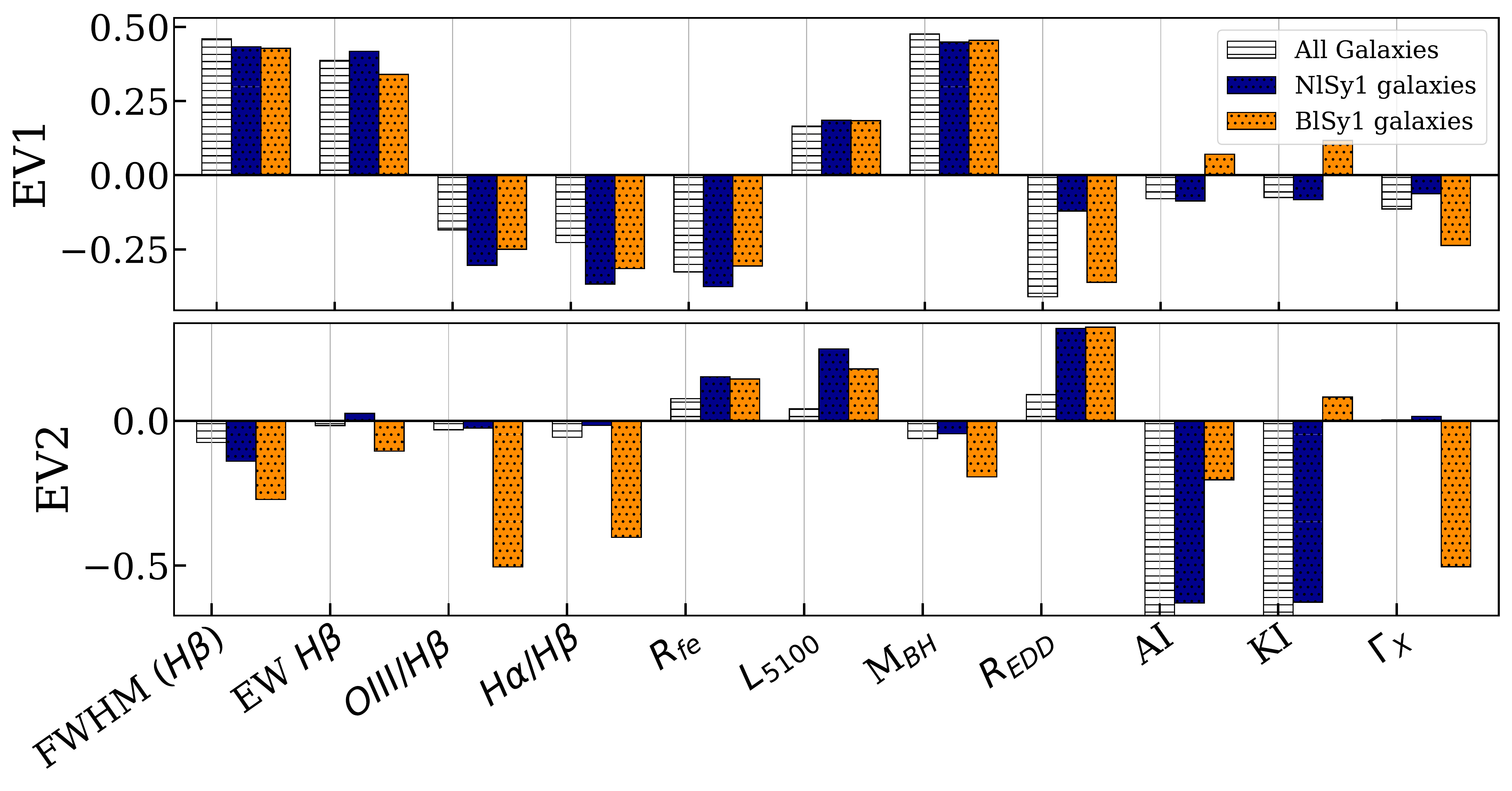}}

\caption{The projection of the 11 physical parameters on the Eigenvector 1 (top) and Eigenvector 2 (bottom) as a result of the PCA performed on the sample of NLSy1 (blue color) and BLSy1 (orange color) galaxies individually and jointly (white color).}
\label{pca_results}
\end{figure*}

The results of the PCA are available in Table \ref{Table3}. The first three Eigenvectors are responsible for almost 66\% of the variation in the sample. The Eigenvector1 is responsible for 37\%  variation in the case of NLSy1 galaxies and 30\% variation in the case of BLSy1 galaxies. The EV1 is dominated by the anti-correlation between the $H\beta$ FWHM and $R_{fe}$ for both the cases, clearly indicating the validity of the relationship obtained in \citet{Boroson1992}. Figure \ref{pca_results} shows the projection of Eigenvector 1 on the various components for both cases. $R_{\textsc{edd}}$  and M$_{BH}$ are the other parameters that dominate the EV1 projections.  While M$_{BH}$ correlates highly with EV1,  $R_{\textsc{edd}}$ anti-correlates weakly with EV1 in the case of NLSy1 galaxies, it anti-correlates strongly with EV1 in the case of BLSy1  galaxies (see Table \ref{table4} for more information). AI shows a weak correlation with EV1 in the case of BLSy1 galaxies, while it shows a negligible correlation with EV1 in the case of NLSy1 galaxies. As only a few AGN show large asymmetry indices, the correlation may not significantly affect the parameter space when analyzing larger AGN populations.  The EV2 is responsible for 19\% (20\%) variations in the case of NLSy1 (BLSy1)  galaxies. The Eigenvector 2 of BLSy1 galaxies is driven by the anti-correlation of $[OIII]/H\beta$ flux ratio,  $\Gamma_X$  and $H\alpha/H\beta$ flux ratio and the correlation of $R_{\textsc{edd}}$. However, in the case of NLSy1 galaxies, the parameters driving the EV2 are AI, KI, and $R_{\textsc{edd}}$ (see Table \ref{Table3}). The Spearman rank correlation coefficients between the EV2 and the parameters yield that $R_{\textsc{edd}}$  highly correlates with EV2 in both cases, meaning that it is a significant driver of EV2. It is interesting to note that the projections of the EV1 and EV2 obtained in this study are different from the ones obtained in \citet{Boroson1992}. While their EV1 shows a high  correlation with $R_{\textsc{edd}}$, in our study, $R_{\textsc{edd}}$ shows anti-correlation with EV1 and a strong correlation with EV2. A similar correlation to our study was obtained in \citet{Jarvela2015}.  Since the Eigenvectors may yield different projections for different datasets \citep{Grupe2004}, it is thus imperative to perform the multi-parameter correlation analysis with different parameters in order to understand the underlying differences.

\begin{table*}

\begin{tabular}{lrrrrrr}
\hline
{\bf Parameter}               & \multicolumn{2}{l}{\bf NLSy1 galaxies} & \multicolumn{2}{l}{\bf BLSy1  galaxies} & \multicolumn{2}{l}{\bf Combined} \\
\hline
                        & EV1              & EV2             & EV1              & EV2             & EV1           & EV2          \\
\hline

FWHM ($H\beta$)   & 0.813  & $-$0.605 & 0.958  & $-$0.674 & 0.924  & $-$0.838 \\
EW $H\beta$             & 0.816  & $-$0.137 & 0.764  & $-$0.352 & 0.749  & $-$0.431 \\
$[OIII]/H\beta$            & $-$0.392 & $-$0.061 & $-$0.184 & $-$0.123 & $-$0.273 & 0.060  \\
$H\alpha/H\beta$        & $-$0.577 & 0.139  & $-$0.552 & $-$0.121 & $-$0.293 & $-$0.082 \\
$R_{fe}$                & $-$0.554 & 0.519  & $-$0.622 & 0.468  & $-$0.693 & 0.700  \\
log($L_{5100}$)              & 0.376  & 0.578  & 0.397  & 0.234  & 0.366  & 0.066  \\
$log(M_{BH}/M_{\odot})$ & 0.871  & $-$0.324 & 0.972  & $-$0.601 & 0.936  & $-$0.777 \\
log($R_{\textsc{edd}}$)    & $-$0.313 & 0.948  & $-$0.780 & 0.806  & $-$0.798 & 0.917  \\
AI                      & 0.058  & $-$0.244 & 0.283  & $-$0.444 & 0.310  & $-$0.420 \\
KI                      &  $-$0.226 & $-$0.032 & 0.187  & 0.132  & $-$0.027 & $-$0.083 \\
$\Gamma_X$            & $-$0.125 & 0.273  & $-$0.200 & 0.117  & $-$0.511 & 0.535 \\

\hline
\end{tabular}
\caption{The Spearman rank correlation coefficient between the different parameters with their corresponding projected values in EV1 and EV2 axis calculated for the sample of NLSy1 galaxies (left), BLSy1  galaxies (middle), and the combined sample (right).}
\label{table4}
\end{table*}

\section{Discussion}
\label{discussion}

To understand the differences in properties of Narrow line Seyfert galaxies compared to the general type-1 Seyfert population, we analyzed the single epoch optical spectrum for a large uniform X-ray selected sample of 144  NLSy1 galaxies and a comparison sample of 117 BLSy1  galaxies. This study is unique because the optical parameter correlation analysis involving emission line shapes for a large sample has been done, keeping the separation between the NLSy1 galaxies and BLSy1  galaxies in mind. Results based on our large sample of NLSy1 and BLSy1  galaxies are consistent with other studies with similar samples \citep[see][]{Grupe2004, Zamfir2010, Xu2012}. The Spearman rank correlations point to higher anti-correlation of the $H\beta$ FWHM with $R_{fe}$ and $R_{\textsc{edd}}$  in the case of NLSy1 galaxies. This provides a strong hint towards the younger age of NLSy1 galaxies. While there is a strong anti-correlation of FWHM with $R_{fe}$, it is slightly weaker in the case of BLSy1  galaxies. This may be because $R_{fe}$ is strongly dependent on the flux of the iron emission line, and thus the emission region of NLSy1 galaxies is richer in iron content as compared to the BLSy1  galaxies \citep{Collin2000}. We find out that the SMBH mass is correlated with FWHM in both the cases and that the NLSy1 galaxies have a higher  $R_{\textsc{edd}}$ than the BLSy1 galaxies. The lower SMBH mass coupled with high accretion rates in the NLSy1 galaxies may indicate that they are in evolving phase as compared to the BLSy1 galaxies. We also note here that the anti-correlation seen between SMBH mass and $R_{\textsc{edd}}$ is influenced by the FWHM, which by definition is systematically lower in NLSy1 as compared to BLSy1 galaxies.


We find out that SMBH mass correlates highly with the FWHM of H$\beta$ emission line, which indicates that the more massive the central SMBH is, the faster the gas in the line emitting region rotates, supporting the virialized motion formalism, despite the significant difference in accretion rates characterized by $R_{\textsc{edd}}$ in both types of galaxies. The $H\alpha$ to $H\beta$ emission line flux ratio is slightly higher in the case of BLSy1 galaxies as compared to the NLSy1 galaxies, which is also evident by its anti-correlations with the FWHM of H$\beta$ emission line. The possible reason could be that the $H\alpha$ and $H\beta$ emission come from different BLR locations. Another possibility could be that the  $H\alpha$  profile is blended with contribution from the NLR emission lines, which are difficult to remove while fitting the $H\alpha$ emission complex. Hence the flux from the NLR might have a contribution.

We have used the line shape parameters, namely the Asymmetry and Kurtosis indices for the H$\beta$ emission line, to understand their influence in the peculiar behavior of the NLSy1 galaxies as compared to the BLSy1 galaxies. The higher number of NLSy1 galaxies showing blue asymmetric profiles adds another peculiar behavior to their nature among the general type-1 Seyfert population. The outflow signatures indicated by blue asymmetric H$\beta$ emission profile has been reported in the past works as well \citep[see][]{Ganchi2019, Wolf2020} and this can be possible due to the accretion disk winds in these kinds of galaxies driven by high accretion rates \citep{Grupe2004, Xu2012}. It is worth noting that there is anti-correlation between $R_{fe}$ and AI in both the types of galaxies, implying that the sources with high $R_{fe}$  tend to show blue asymmetries. However, since the line emitting regions in AGN are still poorly understood, the exact connection between the outflowing components and the ratio of iron flux to the $H\beta$ emission line flux, signified by  $R_{fe}$ is not clear.

The higher value of $R_{fe}$ in NLSy1 is a defining parameter, and this coupled with the high accretion rates may have an influence on these galaxies showing blue asymmetric profiles. As discussed in \citet{Grupe2004}, the NLSy1 galaxies with higher $R_{\textsc{edd}}$ values may induce strong outflows, which might be a possible reason for this phenomenon. It will be interesting to know whether the outflow signatures are limited to only the H$\beta$ emission line in these galaxies or are seen in other emission lines as well. 

We did not find significant correlations of $\Gamma_X$ with the observational and physical parameters in either NLSy1 or BLSy1 galaxies. However, when calculated on the combined population, the correlations of $\Gamma_X$  with $R_{fe}$ and $R_{\textsc{edd}}$  emerged. This may be due to different ranges of the FWHM of H$\beta$ emission line covered in both the populations as has been suggested in \citet{Grupe2004}. The lack of significant correlations of $\Gamma_X$ with the line shape parameters, namely AI and KI, indicates that the emission from the inner regions of the accretion disk does not have a role in driving the emission line asymmetries at least.

\citet{Decarli2008} suggest that the NLSy1 galaxies do not differ from the BLSy1 galaxies generally, and their difference in observed properties can be accounted for by taking the orientation into account. Our results suggest that there are key significant properties that identify the NLSy galaxies as a distinct class in the type-1 AGN population. However, it is also possible that the difference in correlations seen in the NLSy1 and BLSy1 galaxies might be amplified due to the lower range of FWHM values in the former as compared to the latter.

\subsection{ NLSy1 galaxies in the context of Population A and B}

\begin{figure}
\includegraphics[width=9cm,height=7cm]{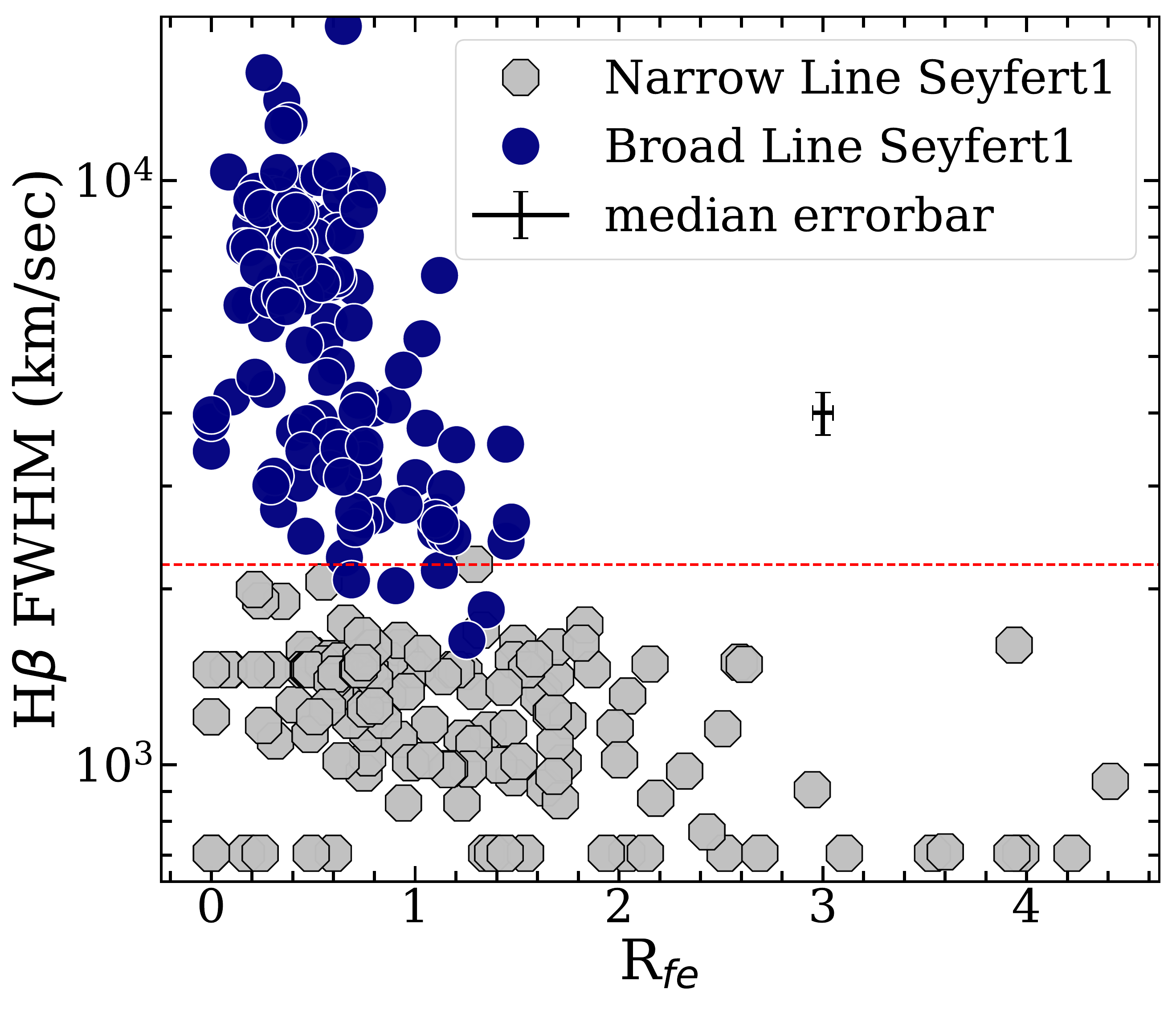}

\caption{The 4DE1 plot for the sample is shown here. The NLSy1 galaxies are shown in grey, while the BLSy1 galaxies are shown in blue. The limit of 2300 km sec$^{-1}$ chosen for differentiating the NLSy1 galaxies from the BLSy1 galaxies is shown as horizontal red dashed line.}
\label{4de1_plot}
\end{figure}

It has been argued that the Population A AGN (FWHM of $H\beta \leq $ 4000 km s$^{-1}$) are well fitted by Lorentzian profiles while the Population B AGN (FWHM of $H\beta >$ 4000 km s$^{-1}$) are well fitted using multiple Gaussian profiles \cite[see][]{Sulentic2009, Marziani2018}. Moreover, high and low accreting sources in this population show significant differences in the parameters. 
Further in \citet{kova2010} a difference in properties is seen at even 3000 km/sec, which is challenging to our understanding of the AGN population and defining the limit of emission line FWHM at which the ensemble properties change. While most of the studies \citep[e.g.,][]{Zamfir2010, Marziani2018} have proposed a difference in the physical properties at $H\beta$ FWHM = 4000 km s$^{-1}$, we tried to see if the differences in physical properties arise even in the case of NLSy1 galaxies when compared to the general Seyfert galaxies population. Naturally, with the NLSy1 galaxies occupying extreme ends in some parameter space, it becomes imperative to understand the properties of NLSy1 galaxies compared to the general type-1 Seyfert population. In \citet{Grupe2004}, and many other studies \citep[e.g., see][]{Xu2012,Ojha2020,Waddell2020} the properties of NLSy1 galaxies have been studied comparatively with BLSy1 galaxies. However, these studies have used a different set of observational and physical parameters from those studied here. Also, it can be noted that in \citet{Grupe2004} and \citet{Xu2012} the sample size was much smaller than the one used here. However, the results of these studies and ours are largely consistent. We also observe that the negative AI values which are prominent in NLSy1 galaxies are not present significantly in the BLSy1 galaxies with H$\beta$ FWHM values lower than 4000 km/sec. Thus, negative AI values may be another parameter that can distinguish the NLSy1 galaxies as a class from the general type 1 Seyfert population. In Figure \ref{4de1_plot}, we show the 4DE1 plot based on this sample. The NLSy galaxies occupy the bottom right space (with low FWHM and larger $R_{fe}$ values). However, we do not find any BLSy1 galaxies with higher $R_{fe}$ values indicating that the NLSy1 galaxies may be occupying a parameter space that separates them from the population of BLSy1 galaxies. 

\subsection{Radio loud NLSy1 galaxies}
 Blue asymmetries have been observed in the narrow [OIII] emission line \citep[see][]{Berton2016, Gaur2019} for both the BLSy1 and NLSy1 galaxies, and in the case of NLSy1 galaxies, this phenomenon has been attributed to the NLR outflows generated due to higher accretion rates. Here, we have observed blue asymmetric H$\beta$ emission profiles predominantly in the NLSy1 galaxies. \citet{Marziani2003} explored the blue outliers in the context of the 4DE1 classification of quasars. In \citet{Berton2016} the influence of radio loudness on the gas dynamics is explored. They find that the likelihood of asymmetric [OIII] emission profiles is higher in the radio-loud AGN than the radio-quiet AGN. They find that the interaction with the relativistic jets in the radio-loud NLSy1 galaxies may give rise to the blue asymmetric profile in the [OIII] emission line. It is indeed possible that a similar mechanism could be responsible for the asymmetry in the H$\beta$ emission line in these kinds of galaxies. \citet{Veeresh2018} presented the analysis for a sample of 11001 NLSy1 from \citep{Rakshit2017} out of which only 498 are detected in the radio wavelength. In our sample of 144 NLSy1 galaxies, we find out that 23 are radio-loud, which may harbor a jet. However, we could not find out the physical properties which differentiate these two classes. We performed a KS test on the AI and KI values for the radio-loud and radio-quiet galaxies. The KS statistic for AI was 0.28 while the p$_{null}$ value was 0.08, which means that the two populations are not statistically different. Based on this, we suggest that the jets in these galaxies may not be drivers behind the asymmetries observed in the H$\beta$ emission profile. In \citep{Marziani2003}, the influence of radio loudness on the broad component of H$\beta$ emission profile is not detected, so our results are in line with this result.

\subsection{Interpretation of PCA Eigenvectors}

From the PCA performed on both the types of galaxies, we find out that the first three Eigenvectors are responsible for approximately 66\% variance in the sample, which is similar to the results obtained in the recent works \citep[e.g., see][]{Grupe2004, Xu2012, Jarvela2015, Wolf2020}. We have included AI and KI, signifying the emission line shapes measured for the $H\beta$ emission line in the PCA for a large sample of NLSy1 and BLSy1 galaxies for the first time. The problem with interpreting the Eigenvectors from the PCA is that every sample may have its own Eigenvector as the physical parameters chosen may be different for different studies \citep{Grupe2004}.  Hence, different parameters may arise as the dominant parameters for individual samples. Through this work, we found out that the EV1 correlates with the FWHM of H$\beta$, SMBH mass, $L_{5100}$  while it anti-correlates with  $R_{fe}$ and $R_{\textsc{edd}}$ and the [OIII]/H$\beta$ and $H\alpha/H\beta$ flux ratios in both the types of galaxies. However, a strong correlation of $R_{\textsc{edd}}$ with EV2 is observed.  The PCA results are different if the two populations are analyzed separately. For instance, the behavior of AI, KI, and $[OIII]/H\beta$ flux ratio with the second Eigenvector in the sample of NLSy1 galaxies differ significantly from the BLSy1 galaxies (see Figure \ref{pca_results}). This implies that the observational and physical parameters responsible for driving the variation in the NLSy1 and BLSy1 galaxies may differ significantly, and the NLSy1 galaxies occupy their own parameter space, which is different from being a subclass of the BLSy1 galaxies. 

\section{Summary and Conclusions}
\label{conclusions}
In this work, we performed a statistical analysis on a large sample of NLSy1 and BLSy1  galaxies to understand the diversity in their parameters. This is a unique sample selected through X-ray observations and having almost similar luminosity and redshift values, which is intended to be a representative sample in parameter space for both types of galaxies. We analyzed the entire sample with direct correlation analysis by calculating the Spearman rank correlation coefficients and then performed a PCA on the entire sample. The main results are summarized as follows:

\begin{enumerate}
    \item Spearman rank correlations yield that the 4DE1 formalism where $R_{fe}$ anti-correlates with the FWHM of H$\beta$ holds true for both the NLSy1 and BLSy1 galaxies despite the obvious differences in their spectral properties and Eddington ratio.
    \item  Higher fraction of NLSy1 galaxies show blue asymmetries (i.e., traces of outflowing gas) compared to the BLSy1  galaxies. This phenomenon may arise due to the presence of higher iron content, characterized by higher $R_{fe}$ values in the line emitting regions of NLSy1 galaxies.
    \item The asymmetry indices correlate weakly with the FWHM of $H\beta$ and anti-correlate with the $R_{fe}$ values indicating that the sources with a lower value of $H\beta$ FWHM and higher $R_{fe}$ values tend to show blue asymmetries. This result also indicates that the asymmetry indices may be a surrogate parameter in 4DE1 formalism; however, the correlation of asymmetry indices and FWHM of $H\beta$  itself is not very strong in this sample.
     \item Higher $R_{\textsc{edd}}$ values for NLSy1 galaxies as compared to the BLSy1  galaxies confirms their peculiar behavior of low mass black holes accreting at higher rates as compared to their Broad-line counterparts.  Moreover, we found strong anti-correlation between  $R_{\textsc{edd}}$ and FWHM of $H\beta$ in both the cases of NLSy1 and the BLSy1 galaxies, which can be interpreted as the NLSy1 galaxies being in the early phases of AGN development and accreting at a faster rate.
    \item The PCA results signify that the first three Eigenvectors can describe around 66\% of variance in both types of galaxies. The EV1 correlates with H$\beta$ FWHM, H$\beta$ EW, SMBH mass, $L_{5100}$ and AI while it anti correlates with $R_{\textsc{edd}}$, $R_{fe}$, and  $H\alpha/H\beta$  flux ratio for the combined population. However, PCA run separately yields different correlations indicating that the NLSy1 occupy a parameter space of their own which is different from that of the BLSy1 galaxies.
\end{enumerate}

While this study is based on around 260 NLSy1 and BLSy1 galaxies, the NLSy1 galaxies occupy only a small fraction of the Seyfert 1 population, but we used a matching sample, which may not be representative of the entire population. Studies with even larger samples will be helpful in the understanding of the general behavior of NLSy1 galaxies compared with the type-1 Seyfert population and provide insights towards understanding the SMBH growth, dynamics of matter in the vicinity of the SMBH as well as the outflows and feeding of such galaxies.

\section*{acknowledgments}
We are thankful to the anonymous referee for providing comments and suggestions, which helped us improve our manuscript considerably. We are thankful to Hengxiao Guo for making the code PyQSOFit public. This work has made use of scikit-learn \citep{scikit-learn}. This research is part of the DST-SERB project under grant no. EMR/2016/001723. VKJ, HC, and AO acknowledge the financial support provided by DST-SERB for this project. VKJ and HC acknowledge IUCAA for providing access to their High-Performance Computing (HPC) facility. VKJ is thankful for the warm hospitality provided by CUHP Dharamshala during his stay.
This work has used data from the Sloan Digital Sky Survey (SDSS).
Funding for the Sloan Digital Sky 
Survey IV has been provided by the 
Alfred P. Sloan Foundation, the U.S. 
Department of Energy Office of 
Science, and the Participating 
Institutions. SDSS-IV acknowledges support and 
resources from the Center for High 
Performance Computing  at the 
University of Utah. The SDSS 
website is www.sdss.org.

SDSS-IV is managed by the 
Astrophysical Research Consortium 
for the Participating Institutions 
of the SDSS Collaboration including 
the Brazilian Participation Group, 
the Carnegie Institution for Science, 
Carnegie Mellon University, Center for 
Astrophysics | Harvard \& 
Smithsonian, the Chilean Participation 
Group, the French Participation Group, 
Instituto de Astrof\'isica de 
Canarias, The Johns Hopkins 
University, Kavli Institute for the 
Physics and Mathematics of the 
Universe (IPMU) / University of 
Tokyo, the Korean Participation Group, 
Lawrence Berkeley National Laboratory, 
Leibniz Institut f\"ur Astrophysik 
Potsdam (AIP),  Max-Planck-Institut 
f\"ur Astronomie (MPIA Heidelberg), 
Max-Planck-Institut f\"ur 
Astrophysik (MPA Garching), 
Max-Planck-Institut f\"ur 
Extraterrestrische Physik (MPE), 
National Astronomical Observatories of 
China, New Mexico State University, 
New York University, University of 
Notre Dame, Observat\'ario 
Nacional / MCTI, The Ohio State 
University, Pennsylvania State 
University, Shanghai 
Astronomical Observatory, United 
Kingdom Participation Group, 
Universidad Nacional Aut\'onoma 
de M\'exico, University of Arizona, 
University of Colorado Boulder, 
University of Oxford, University of 
Portsmouth, University of Utah, 
University of Virginia, University 
of Washington, University of 
Wisconsin, Vanderbilt University, 
and Yale University.

\section*{Data Availability}

The data underlying this article is available publicly in the SDSS-DR16, XMM-NEWTON, and ROSAT databases, respectively.

\bibliographystyle{highlights}
\bibliography{main}

\label{lastpage}

\end{document}